%% file: paper.tex
\pdfoutput=1
\documentclass[journal]{vgtc}                

\usepackage[utf8]{inputenc}

\ifpdf
  \pdfoutput=1\relax                   
  \usepackage{graphicx}                
  \DeclareGraphicsExtensions{.pdf,.png,.jpg,.jpeg} 
\else
  \usepackage{graphicx}                
  \DeclareGraphicsExtensions{.eps}     
\fi%

\graphicspath{{figures/}{pictures/}{images/}{./}} 

\usepackage{microtype}                 
\PassOptionsToPackage{warn}{textcomp}  
\usepackage{textcomp}                  
\usepackage{mathptmx}                  
\usepackage{times}                     
\usepackage{cite}                      
\usepackage{tabu}                      
\usepackage{booktabs}                  

\onlineid{0}

\vgtccategory{Research}
\vgtcpapertype{algorithm/technique}

\usepackage{balance}
\usepackage{xspace}
\usepackage{color}
\usepackage{enumitem}
\usepackage{wrapfig}
\usepackage{balance}
\usepackage{xspace}
\usepackage[table,xcdraw]{xcolor}
\definecolor{LightGray}{gray}{0.95}
\definecolor{OliveGreen}{rgb}{0,0.6,0}

\newcommand{\ie}{\emph{i.e.}\xspace}
\newcommand{\eg}{\emph{e.g.}\xspace}

\newcommand{\myparagraph}[1]{\smallskip\noindent\textbf{#1}\xspace}
\newcommand{\vect}[1]{\mathbf{#1}}

\newcommand{\figref}[1]{\hyperref[#1]{\textbf{Figure~\ref*{#1}}}}
\newcommand{\figsref}[1]{\hyperref[#1]{Figures~\ref*{#1}}}
\newcommand{\tabref}[1]{\hyperref[#1]{Tab.~\ref*{#1}}}
\newcommand{\secref}[1]{\hyperref[#1]{Sec.~\ref*{#1}}}
\newcommand{\algref}[1]{\hyperref[#1]{Alg.~\ref{#1}}}
\newcommand{\phrase}[1]{``\texttt{#1}''}
\newcommand{\aca}{\texttt{ACA}\xspace}
\newcommand{\web}{\texttt{WEB}\xspace}

\title{ChartText: Linking Text with Charts in Documents}

\author{Joao Pinheiro, and Jorge Poco}
\authorfooter{
\item Joao Pinheiro is with Universidad Catolica San Pablo. E-mail: jaoxvalen@gmail.com.
\item Jorge Poco is with Funda\c{c}\~ao Getulio Vargas, Brazil. E-mail: jorge.poco@fgv.br.
}

\shortauthortitle{Biv \MakeLowercase{\textit{et al.}}: Global Illumination for Fun and Profit}

\abstract{
Recent works show that interactive documents connecting text with visualizations facilitate reading comprehension. However, creating this type of content requires specialized knowledge. 
We present ChartText, a method that links text with visualizations in this work. Our approach supports documents that include bar charts, line charts, and scatter plots. ChartText receives the visual encoding of the visualization and its associated text as input. It then performs the linking in two stages: The matching stage creates individual links relating simple phrases between the text and the chart. Then, it combines the individual links according to the visual channels in the grouping stage, building more meaningful connections.
We use two datasets to design and evaluate our method; the first comes from web documents (24 bar charts and texts) and the second from academic documents (25 bar charts, 25 line charts, and 25 scatter plots with their texts). Our experiments show that our method obtains F1 scores of 0.50 and 0.66 on both datasets. We can also use a semi-automatic approach correcting individual links; in this case, the scores rise to 0.68 and 0.84, respectively.
To show the usefulness of our technique, we implement two proofs of concept. We create interactive documents using graphic overlays in the first one, facilitating the reading experience. We use voice instead of text to annotate charts in real-time in the second. For example, in a videoconference, our technique can automatically annotate a chart following the presenter's description.
}

\CCScatlist{ 
 \CCScat{K.6.1}{Management of Computing and Information Systems}%
{Project and People Management}{Life Cycle};
 \CCScat{K.7.m}{The Computing Profession}{Miscellaneous}{Ethics}
}

\input{0-figures}

\vgtcinsertpkg

\begin{document}


\maketitle

\input{1-introduction}

\input{2-related}
\input{3-data}
\input{4-technique}
\input{5-evaluation}

\input{6-apps}

\input{7-limitations}
\input{8-conclusion}



\bibliographystyle{abbrv-doi}

\bibliography{paper}
\balance
 
\end{document}

%% file: 0-figures.tex
\newcommand{\figMotivation}{
\begin{figure}[t]
  \centering
  \includegraphics[width=\columnwidth]{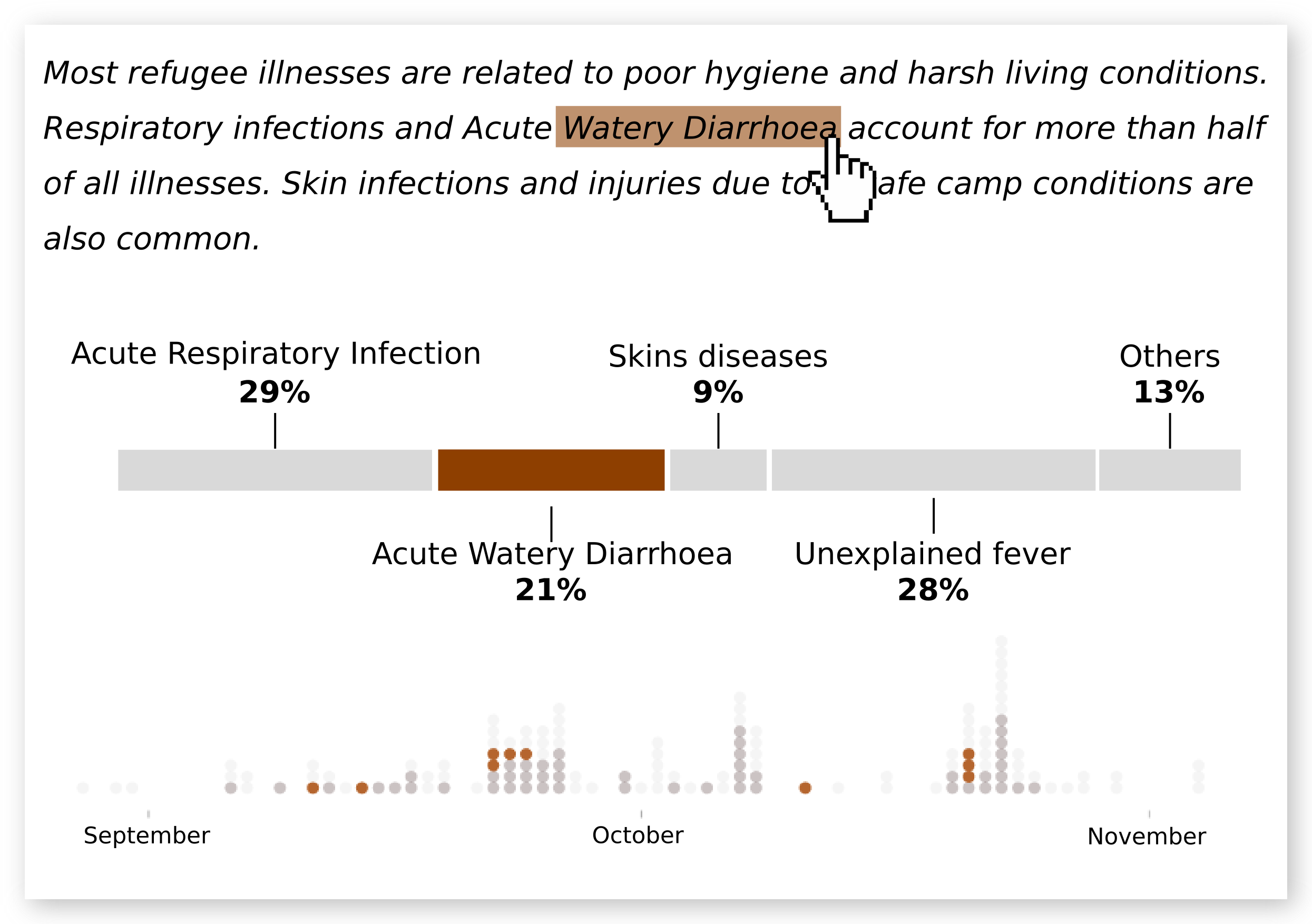}
  \caption{Screenshot of an interactive document taken from the news website Reuters Graphics~\cite{reutergraphics}. It shows how the links between the text and the chart are highlighted to provide a visual guide to the reader.}
  \label{fig:journalism}
\end{figure}

}

\newcommand{\figVisualEncoding}{
\begin{figure}[t!]
  \centering
  \includegraphics[width=\columnwidth]{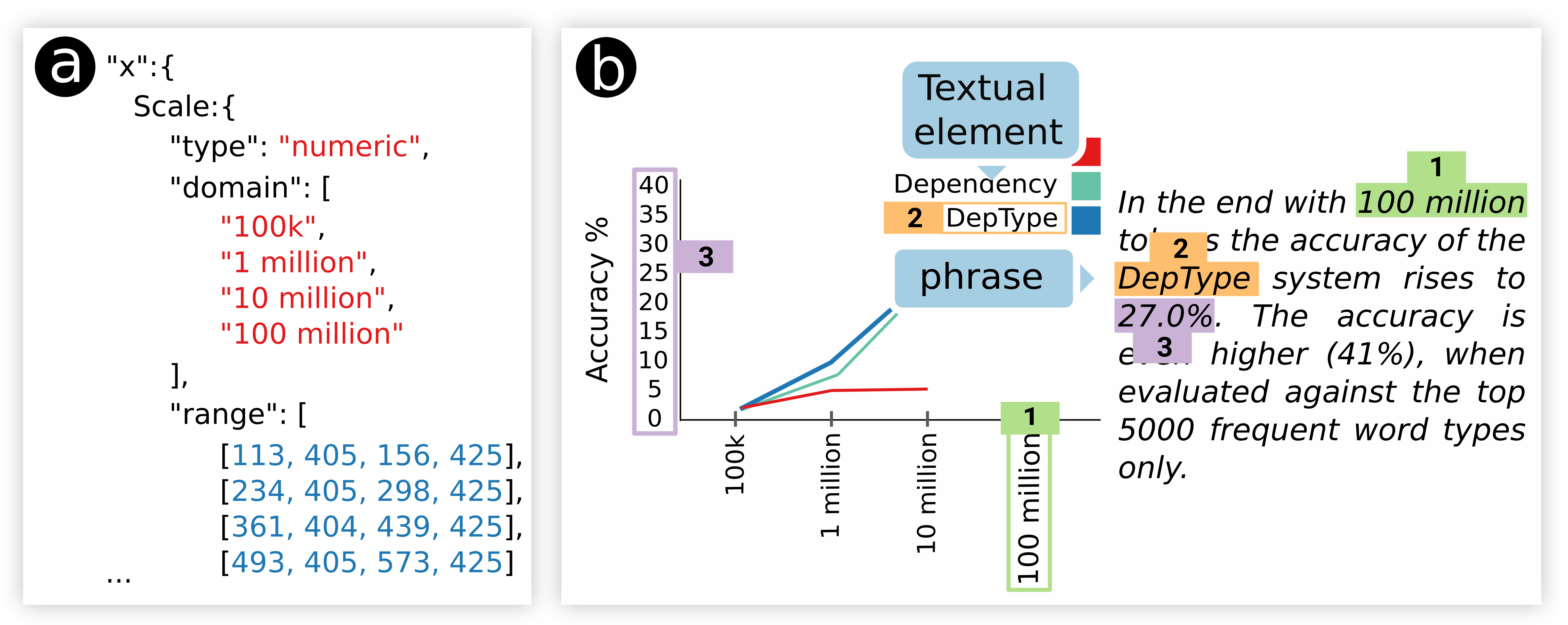}
  \caption{(a) Example of a visual encoding that describes the x-axis of the line chart. 
(b) Example of individual links. The phrases in the paragraph are related to the texts in the chart. We use the same color and numbers to represent the individual links. These individual links are later combined to form grouped links.}
  \label{fig:link_types}
\end{figure}
}

\newcommand{\figTree}{
\begin{figure}[t]
  \centering
  \includegraphics[width=\columnwidth]{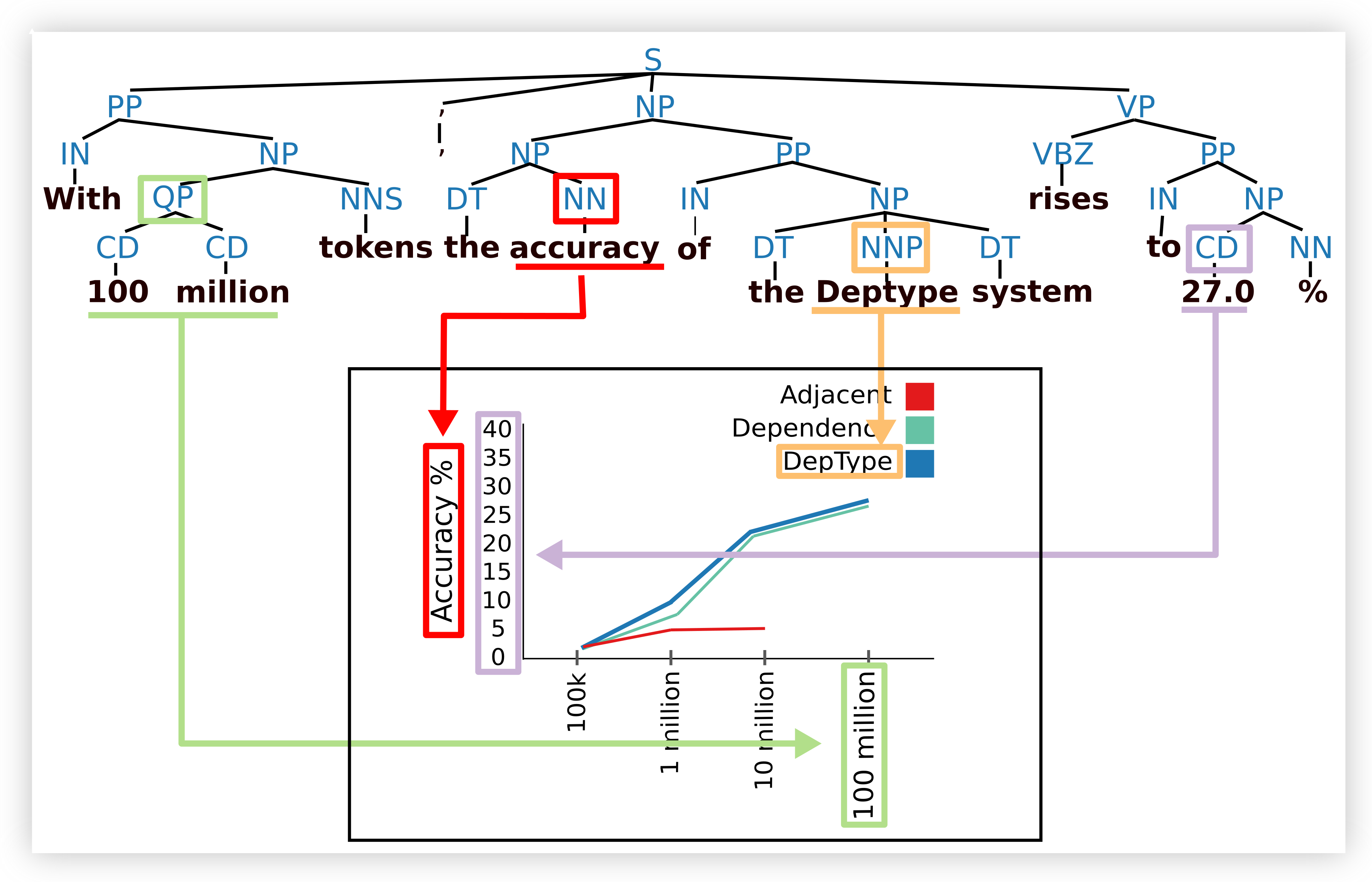}
  \caption{Constituency tree for the paragraph related to the line chart on the bottom.}
  \label{fig:tree}
\end{figure}
}

\newcommand{\figUniqueWord}{
\begin{figure}
  \centering
  \includegraphics[width=\columnwidth]{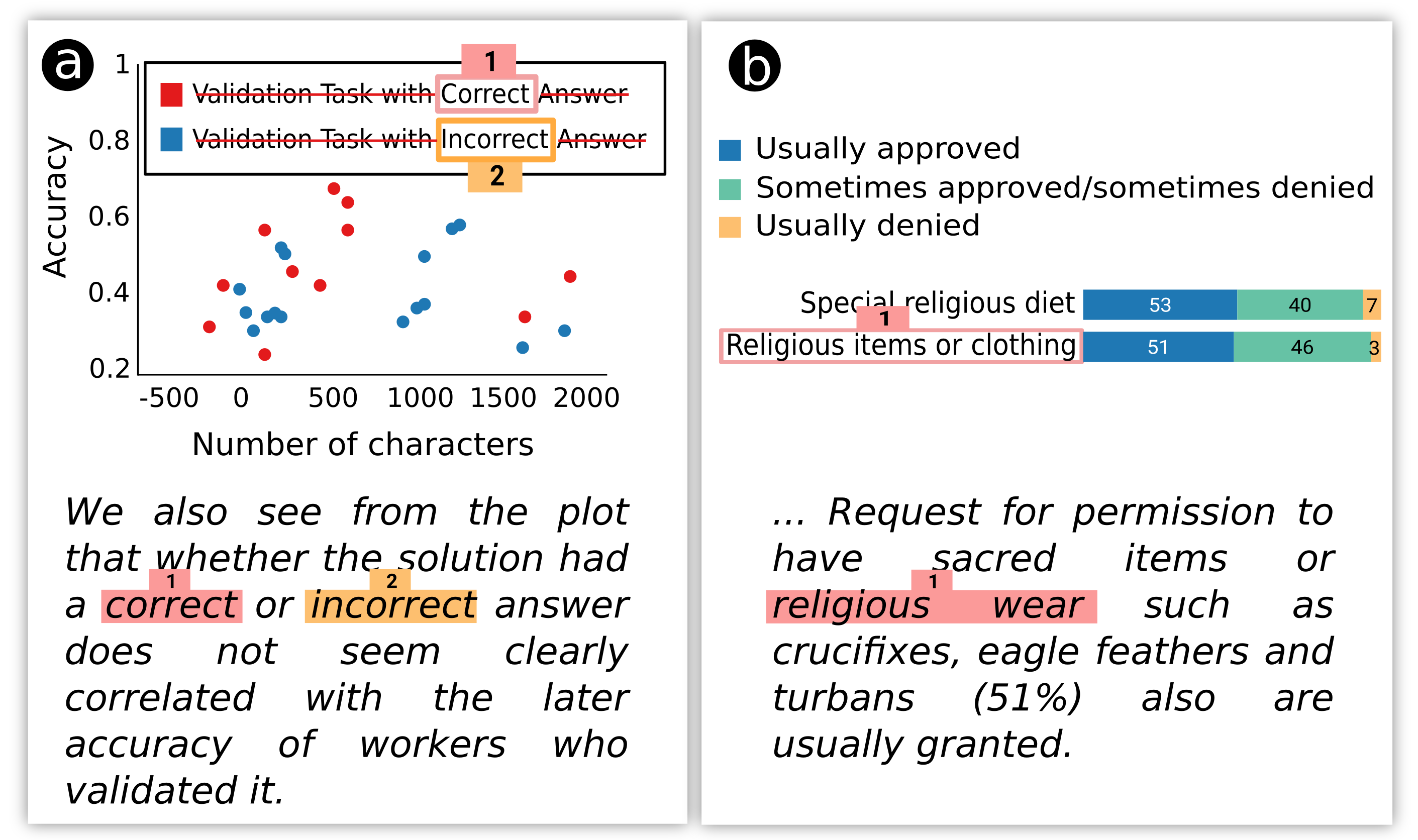}
  \caption{(a) Example of matching with the unique words strategy. (b) Example of matching with the semantic comparison strategy.}
  \label{fig:uniqueword}
\end{figure}
}

\newcommand{\figBadNumeric}{
\begin{figure}[t]
  \centering
  \includegraphics[width=\columnwidth]{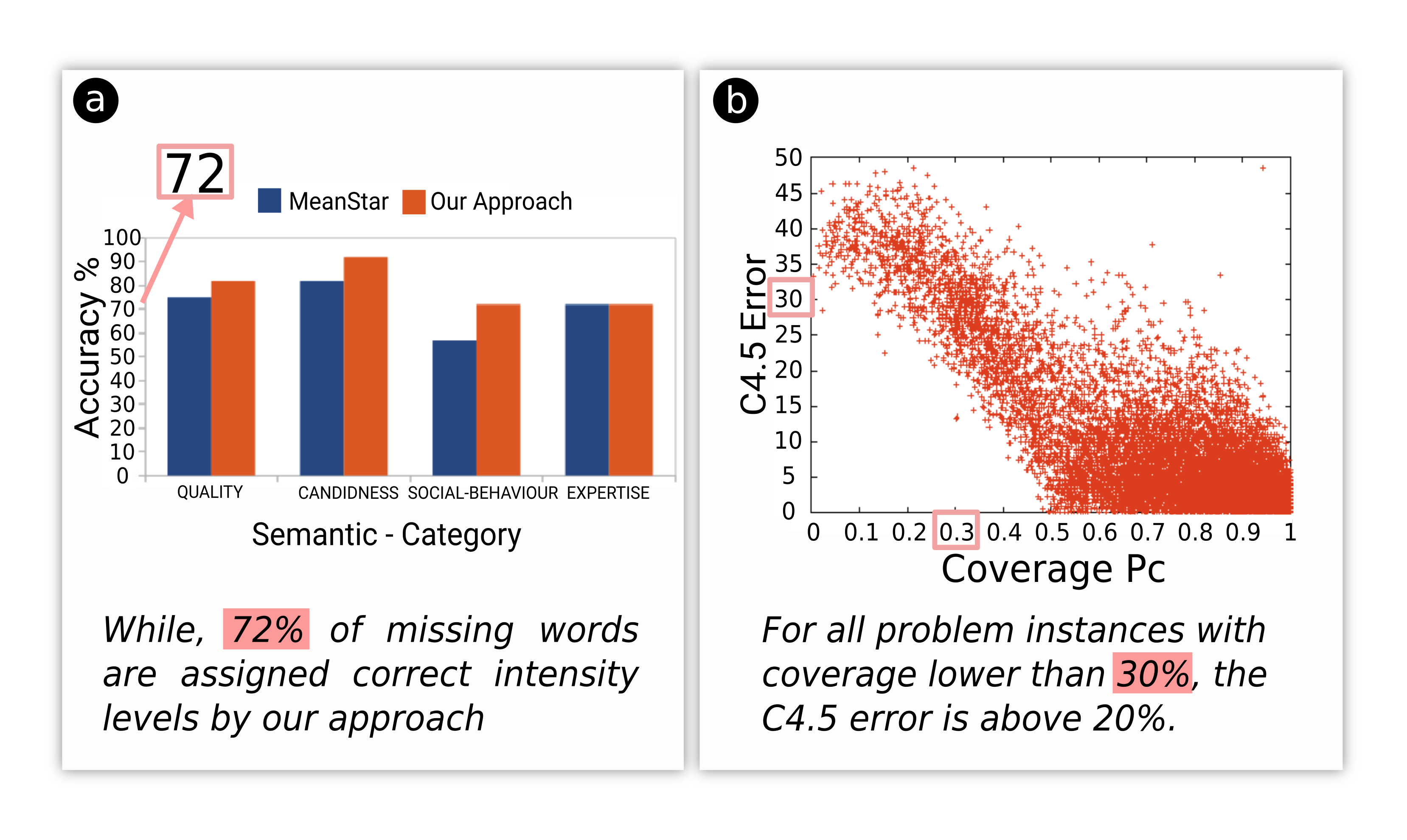}
  \caption{Initial errors in numerical comparison: 
(a) Assignment of the number (\phrase{72}) to an incorrect axis. 
(b) Number \phrase{30\%} could be associated to both numerical axes.}
  \label{fig:badnumeric}
\end{figure}
}

\newcommand{\figRefinement}{
\begin{figure}[t]
  \centering
  \includegraphics[width=\columnwidth]{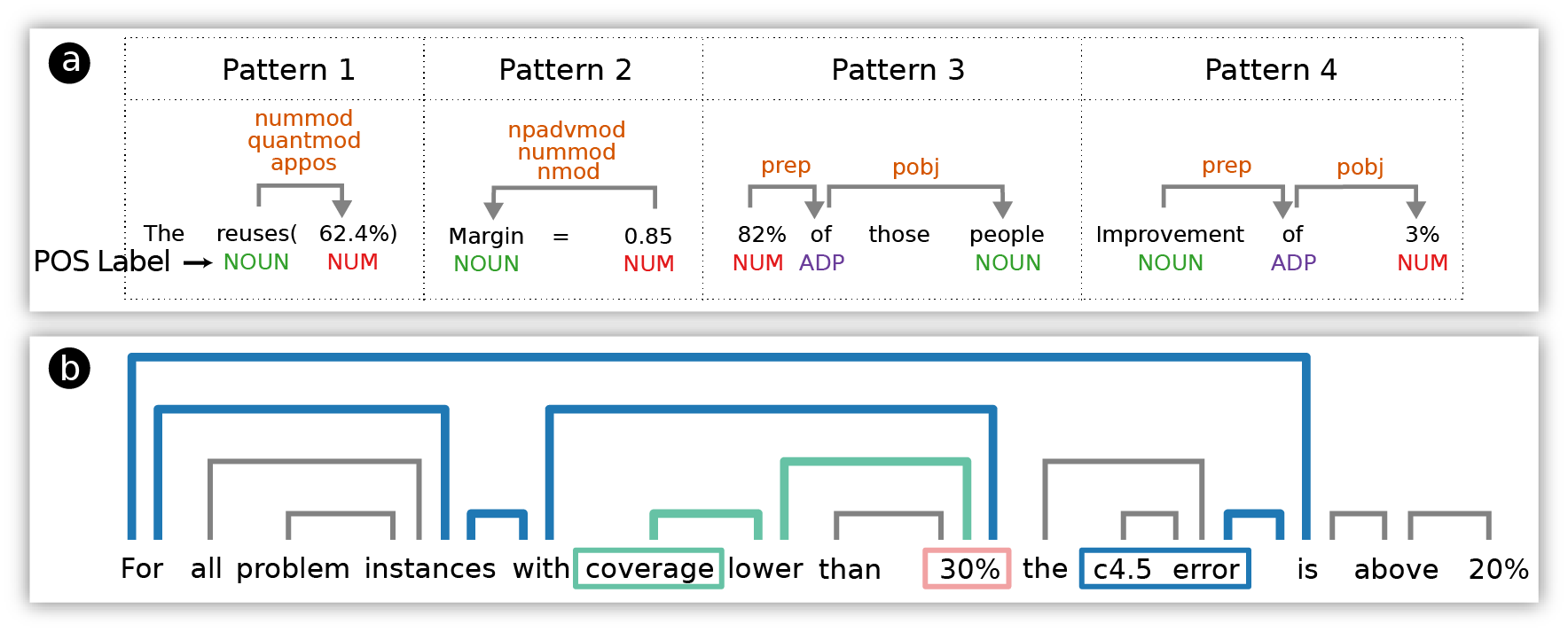}
  \caption{(a) Patterns are used to identify the attribute of a numerical phrase.
  (b) Ambiguity problem, using the syntactic dependency tree, we find the distances between the numerical phrase (\phrase{30\%}) and the phrases associated with the axes titles (\phrase{coverage} and \phrase{C4.5 error}), we keep the phrase with the shortest distance.}
  \label{fig:refnum1}
\end{figure}
}

\newcommand{\figGroups}{
\begin{figure}[t]
  \centering
  \includegraphics[width=\columnwidth]{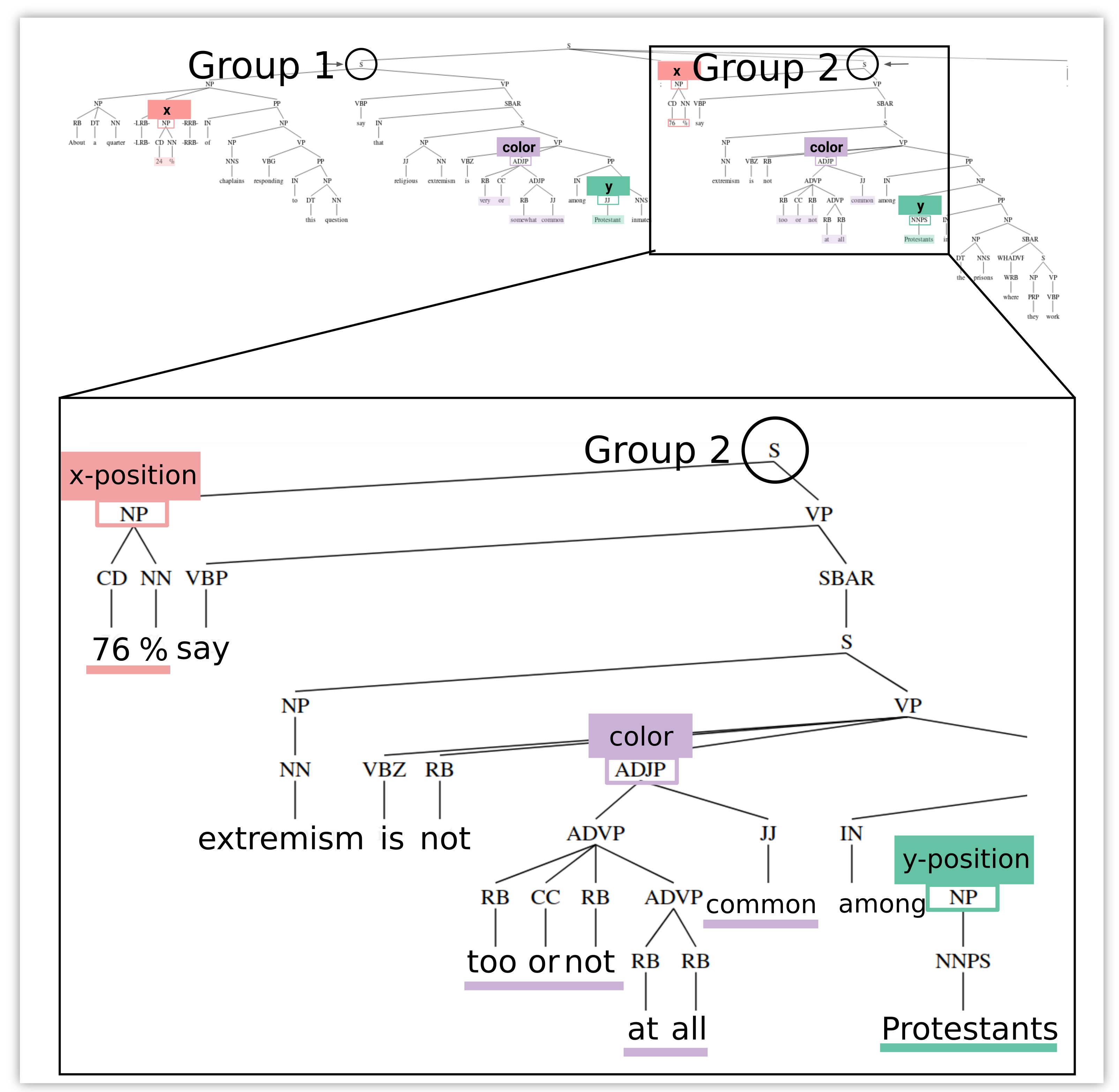}
  \caption{Constituency tree used in the grouping stage.  In the amplified region, we see a node that forms a grouped link (group 2) because it holds three different visual channels.}
  \label{fig:groups}
\end{figure}
}

\newcommand{\figResult}{
\begin{figure*}[t]
  \centering
  \includegraphics[width=\textwidth]{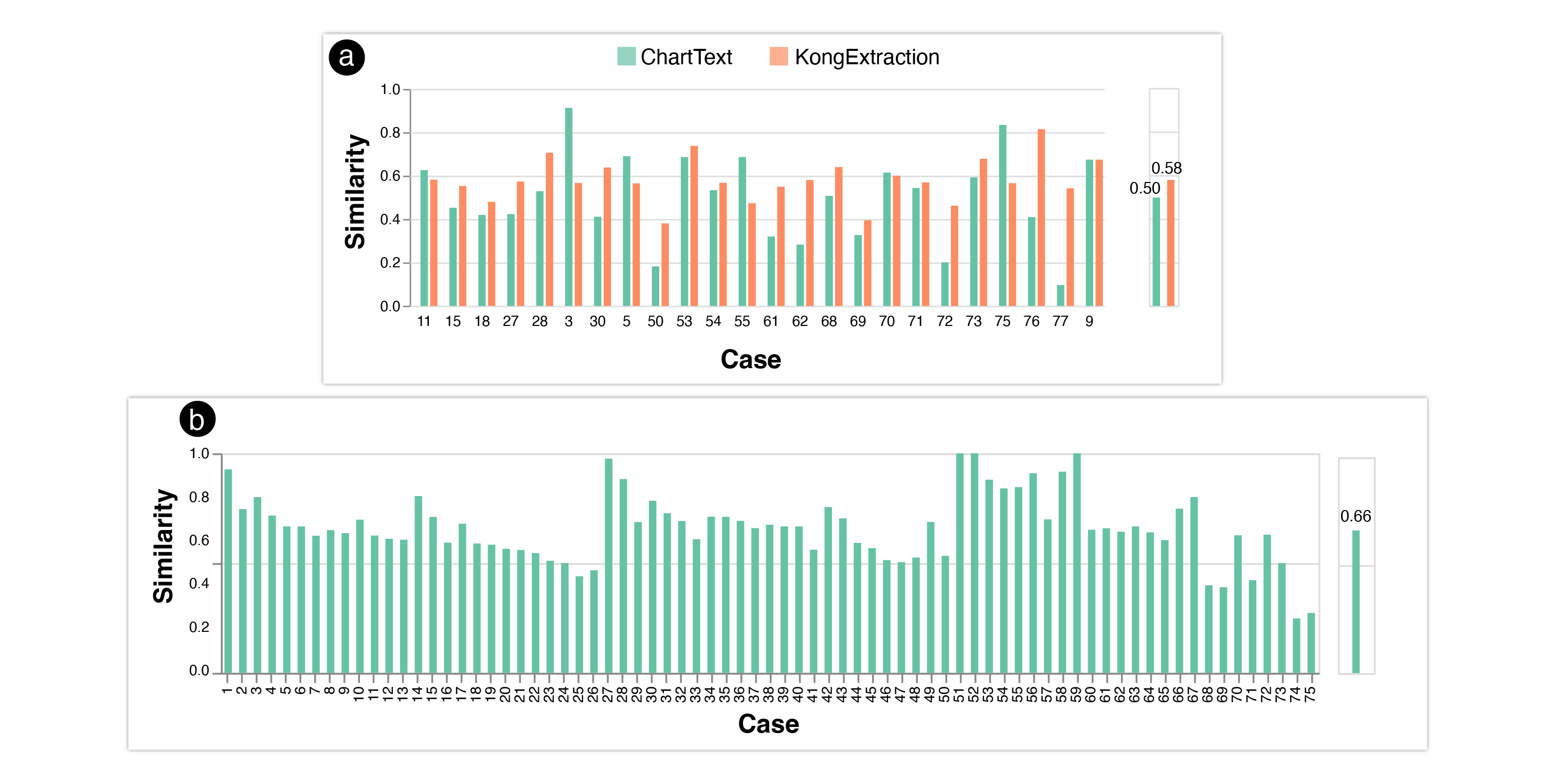}
  \caption{(a) Similarity for each case \web datasets, obtained by our method and by the method proposed by Kong~et~al.\cite{kong2014extracting}. (b) Similarity for each case of the \aca dataset. Each bar represents the similarity with respect to the GS set.}
  \label{fig:result_1}
\end{figure*}
}

\newcommand{\figSimilitud}{
\begin{figure}[t]
  \centering
  \includegraphics[width=\columnwidth]{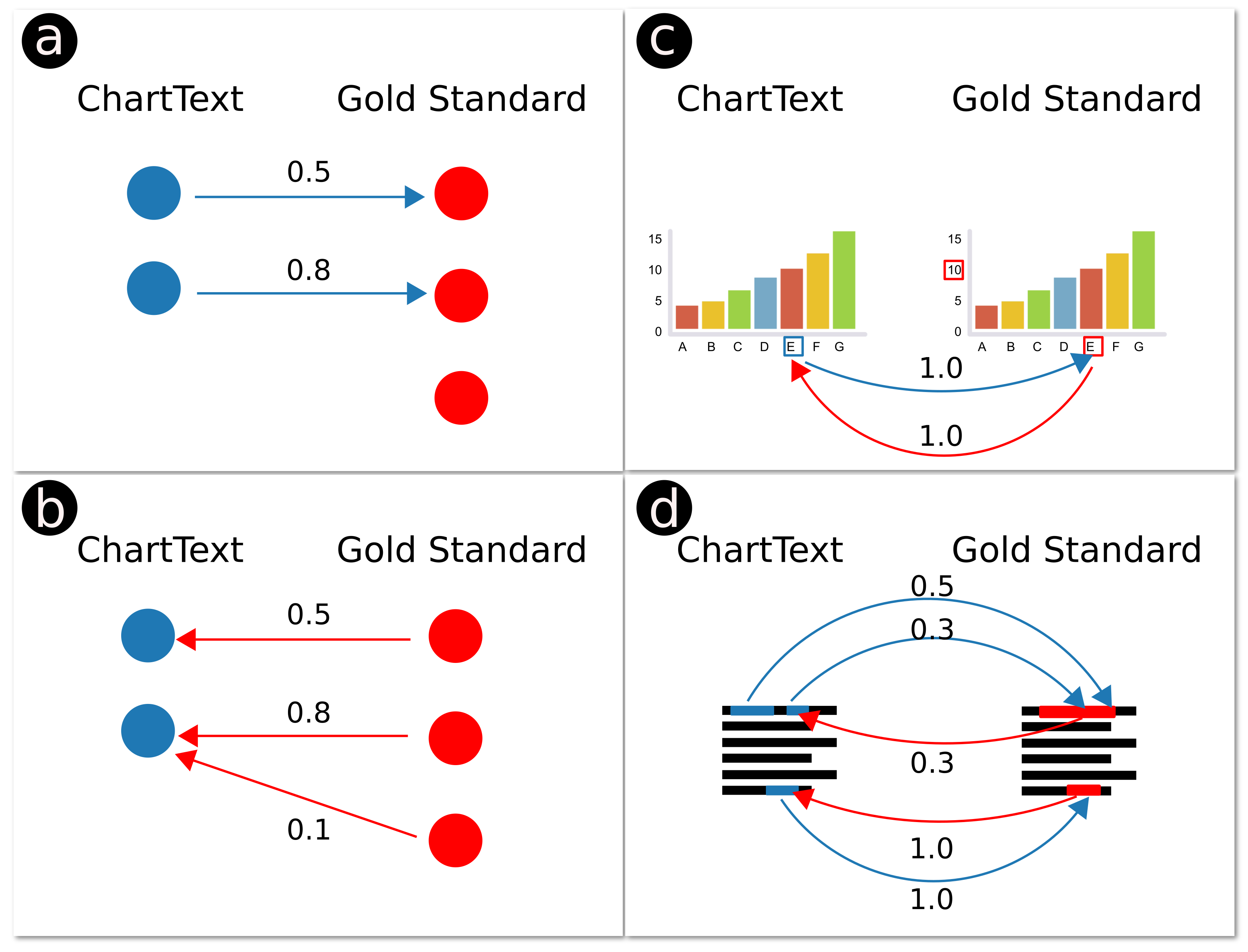}
  \caption{Precision and Recall calculation in the different stages of the metrics. To calculate the Precision, we connect the ChartText results to the gold standard set (GS) one by one, sum the similarities among the connected pairs, and divide by the number of connections. Similar reasoning is used to calculate the recall, but we go from the GS to the ChartText results. (a) Precision calculation among grouped links. (b) Recall calculation among grouped links. (c) Precision and Recall calculation among textual elements. (d) Precision and Recall calculation among phrases.}
  \label{fig:similitud}
\end{figure}
}

\newcommand{\figOverlays}{
\begin{figure*}[t]
  \centering
  \includegraphics[width=\textwidth]{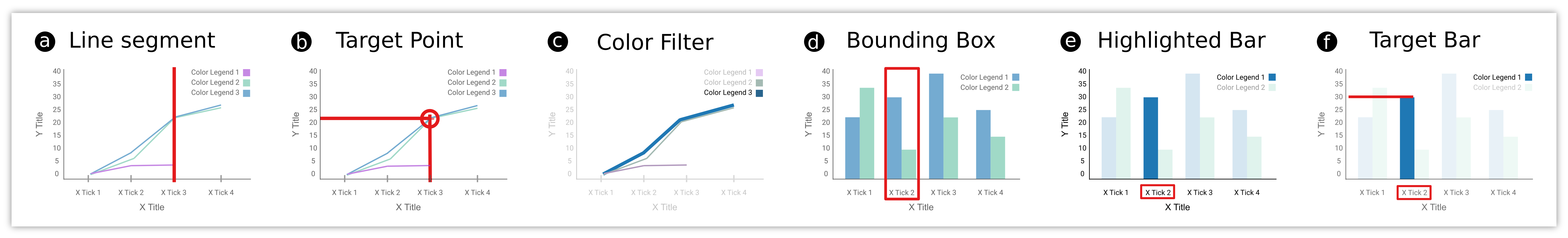}
  \caption{Different types of overlays that are possible to generate with our automatic method. Each type depends on the information associated with the visual channel of each link that makes up the grouped link.}
  \label{fig:lineoverlays}
\end{figure*}
}

\newcommand{\figOverlayApp}{
\begin{figure}[t]
  \centering
  \includegraphics[width=\columnwidth]{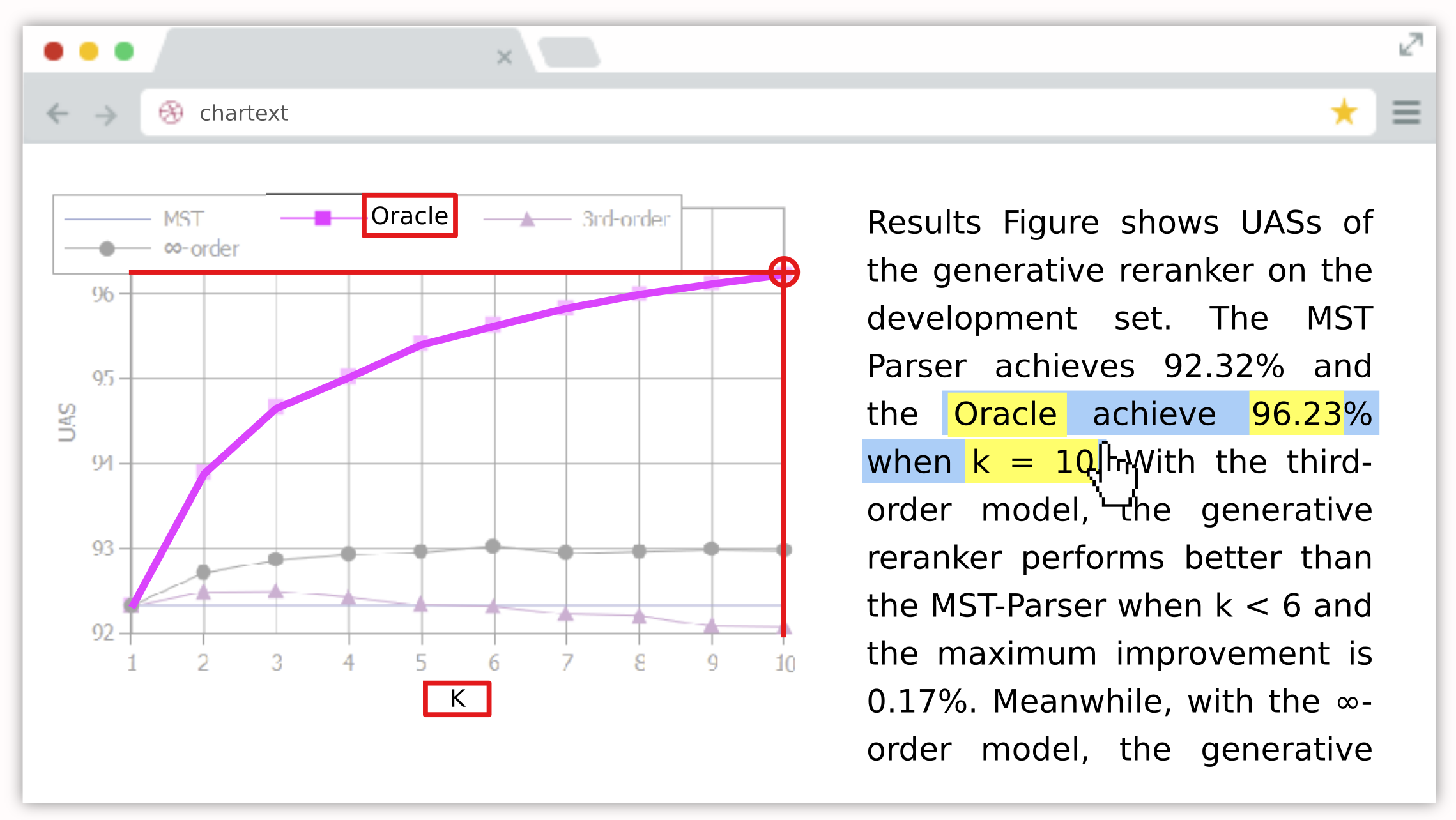}
  \caption{Web application to visualize the overlays. On the left is shown the chart, and on the right, the text describing it. When the user selects a part of the text, the system automatically generates the overlays that are shown on the chart.}
  \label{fig:overlay_example}
\end{figure}
}

\newcommand{\figVoiceOverlaysOne}{
\begin{figure*}[t]
  \centering
  \includegraphics[width=\textwidth]{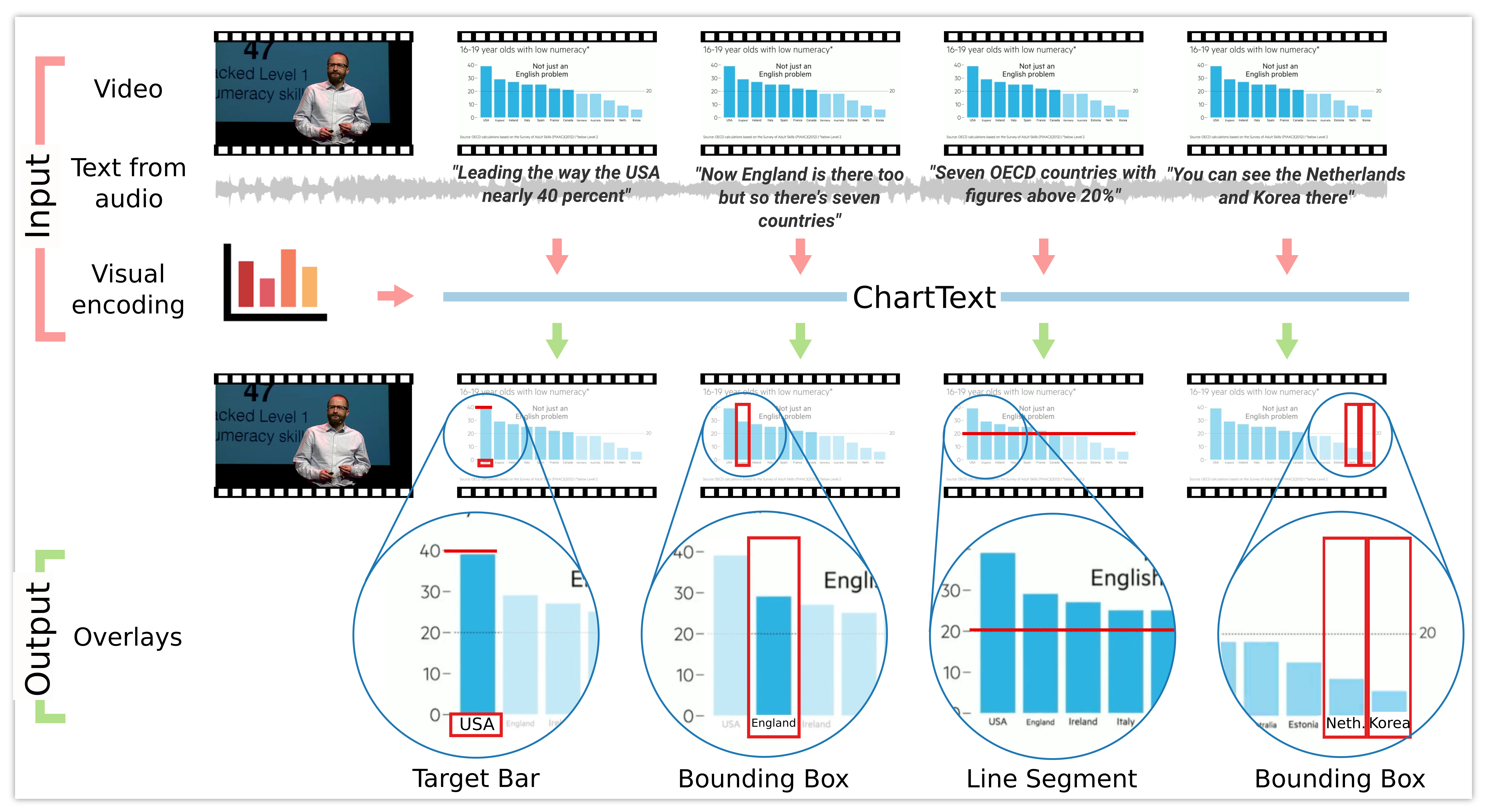}
  \caption{The application uses our proposed method to convert voice into visual guides or overlays on the chart. First, the application receives a video where a presenter describes a chart to the audience. The video’s voice is extracted and converted into text using a speech-to-text tool. The application also receives the visual encoding of the chart in the video as input. Then, the texts extracted from the audio are processed by ChartText, generating as output the overlays shown in the lower part of the figure.}
  \label{fig:app2}
\end{figure*}
}

\newcommand{\figVoiceOverlaysTwo}{
\begin{figure}[t]
  \centering
  \includegraphics[width=\columnwidth]{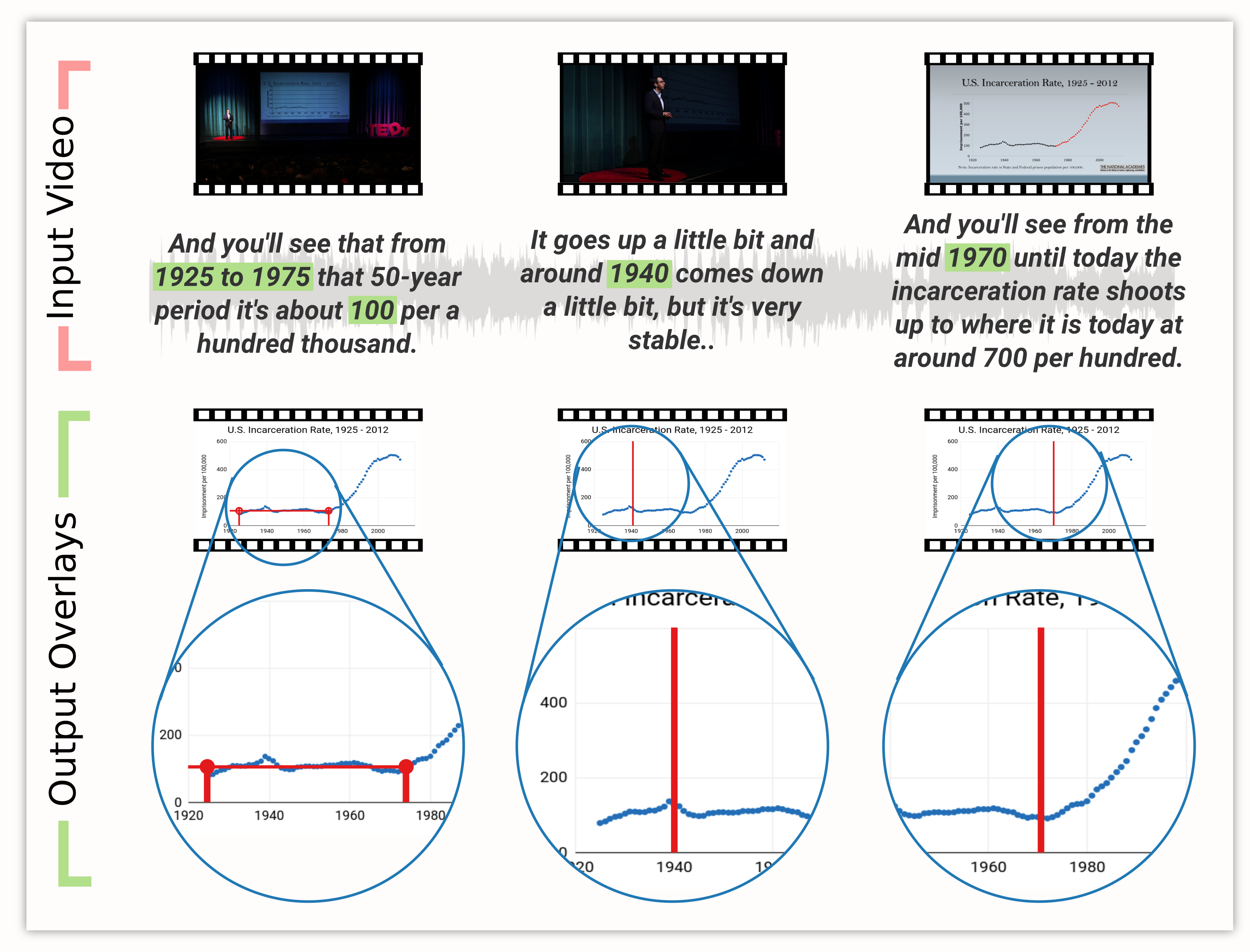}
  \caption{Result of the Voice2Overlays application to convert a presenter’s voice into overlays in real-time. At the top are shown the frames and the extracted phrases from the original video used as input. At the bottom are shown the overlays that our method obtained as output.}
  \label{fig:v2o_case2}
\end{figure}
}

\newcommand{\figErroneous}{
\begin{figure*}[t]
  \centering
  \includegraphics[width=\textwidth]{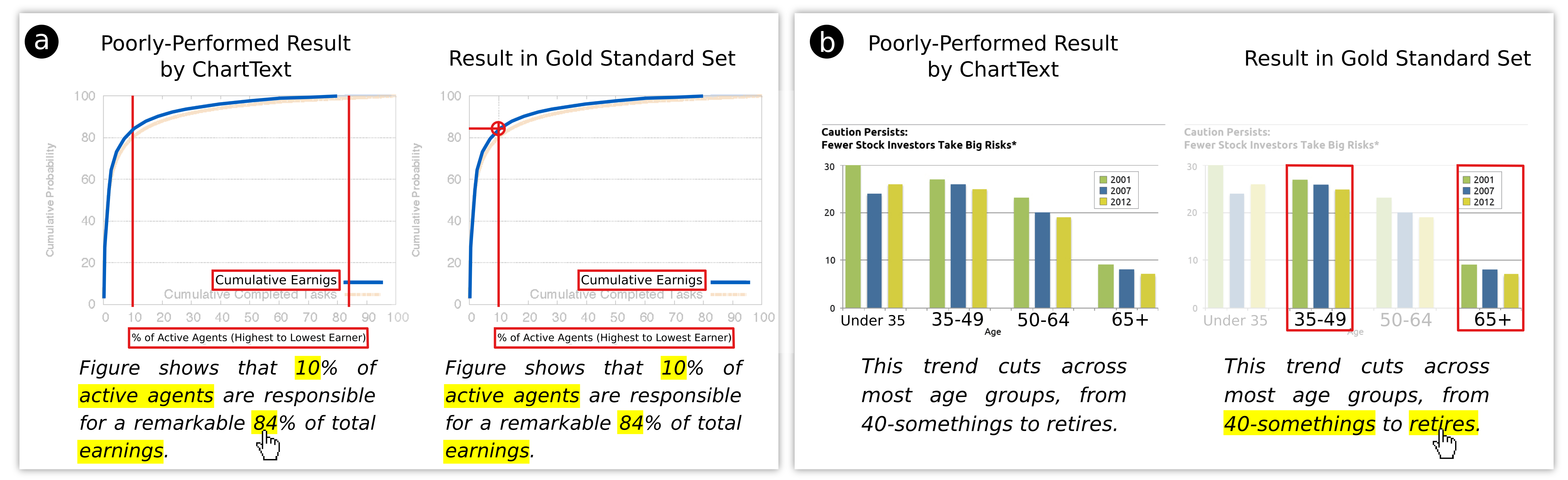}
  \caption{ChartText poorly-performed results. On the left and right side of each case are shown the overlays obtained using ChartText and the gold standard set, respectively. (a) ChartText fails to link the phrase \phrase{84} to the x-axis of the chart. (b) ChartText does not find any link between the text and the chart.}
  \label{fig:erroneous}
\end{figure*}
}

\newcommand{\figPhasesEval}{
\begin{figure}[t]
  \centering
  \includegraphics[width=\columnwidth]{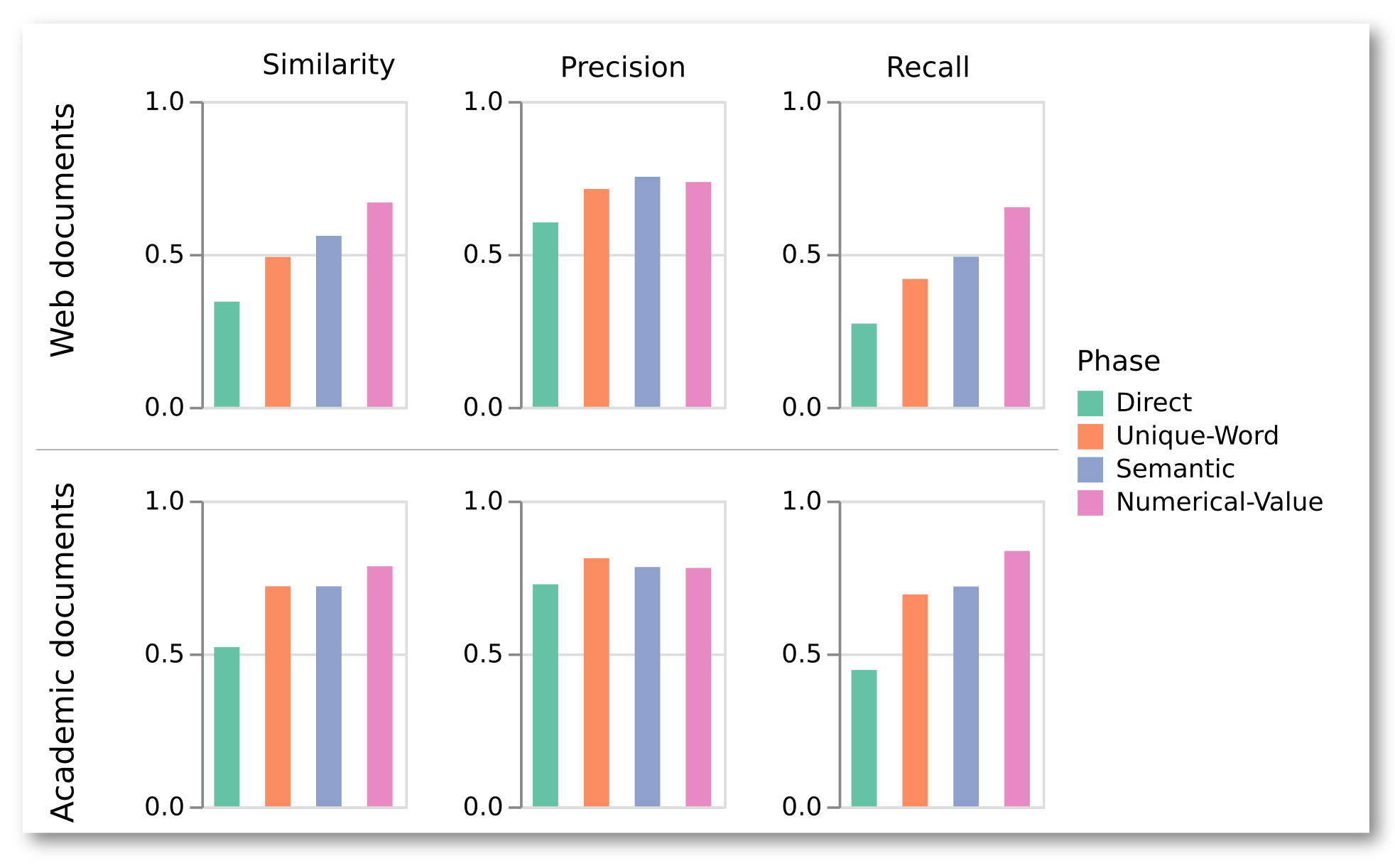}
  \caption{Scores for each phase in our pipeline; we can see the similarity (score F1), precision and recall for each phase in both data sets.}
  \label{fig:phaseseval}
\end{figure}
}

\newcommand{\figSemiauto}{
\begin{figure}[t]
  \centering
  \includegraphics[width=\columnwidth]{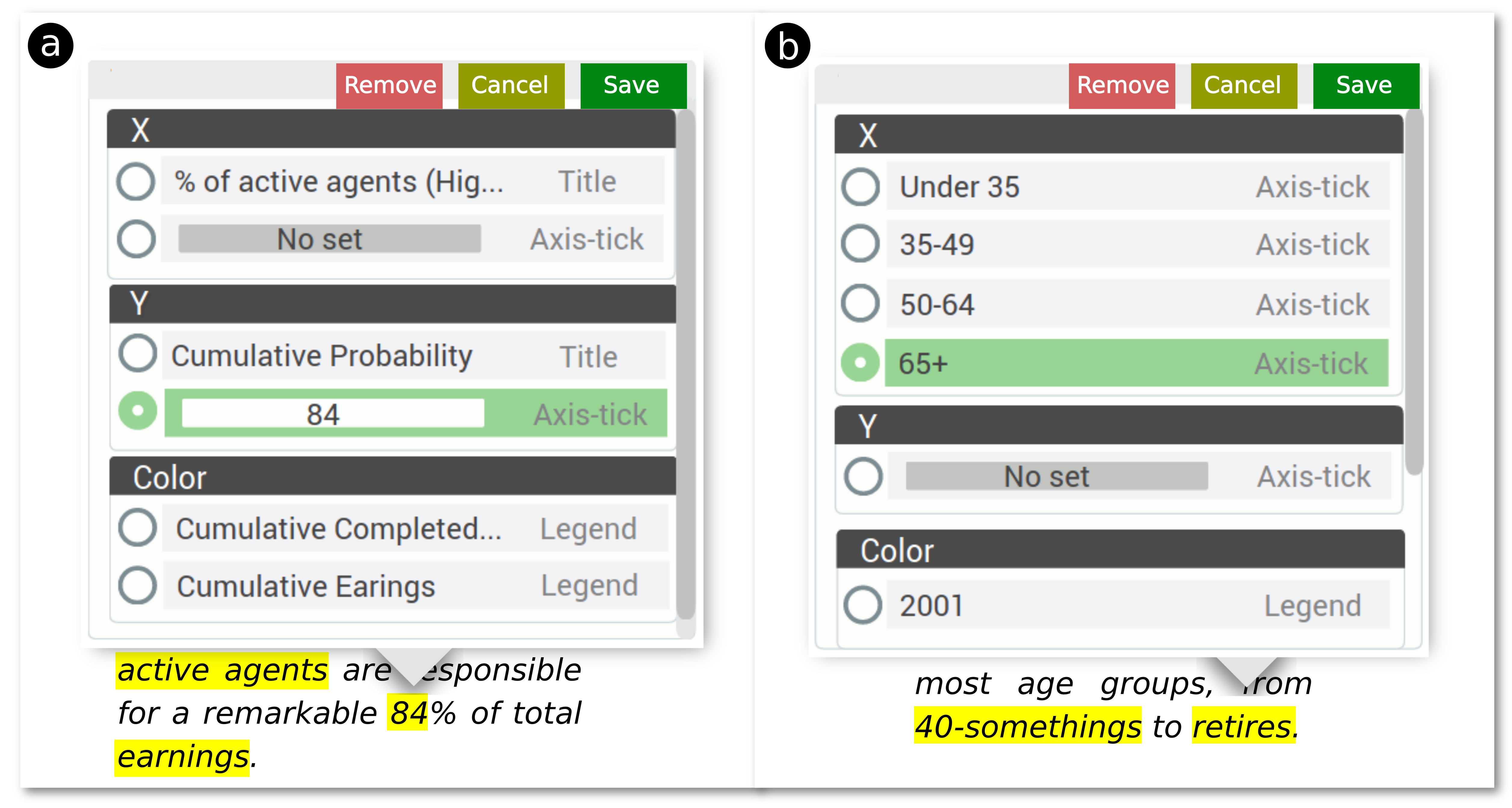}
  \caption{Examples of the use of our semi-automatic component to correct individual links in our overlay web tool.}
  \label{fig:semiauto}
\end{figure}
}

%% file: 1-introduction.tex
\section{Introduction}
\label{sec:introduction}

Visualizations are relevant in the reading comprehension process because they help understand data more naturally. Many content sources, such as scientific papers and news websites, present the information using charts, diagrams, or tables that complement the text to improve understanding of concepts and ideas~\cite{echeverria2009external}. However, to read this kind of document, we must split our attention between elements and mentally integrate the story, requiring more cognitive work~\cite{sweller2011split}.
In response to this problem, some content creators incorporate interactive charts in their documents that serve as a visual guide for the reader~\cite{segel2011storytelling}. In~\figref{fig:journalism}, we can see an example where the user hovers a text, and the chart highlights the associated region. This type of interactivity facilitates the exploration of the document content.

For instance, imagine a news website writer who wants to embed an existing graphic that complements a text presented in an article to generate an interactive experience that guides the reader to a quick understanding of the text-graphic document.
Generating interactive content is a tedious task, even with programming knowledge. Also, the content creator must manually annotate the links between the text and the graphic. Moreover, the writer does not always have access to edit an existing chart or the underlying data to redesign the chart for this type of content. Having a tool that reduces the effort in creating interactive content from text-graphics is the primary motivation for this research.
To build a tool that automatically generates interactive text-chart documents, it is necessary to combine the three main tasks studied in recent years:
(i) it identifies the chart in a document and maps it with the text that references it~\cite{pdffigures2, siegel2016figureseer}
(ii) obtain the underlying data that makes up the chart (\ie, identify the type of chart, get the texts that appear on the chart, identify the visual channels)~\cite{savva2011revision,poco2017reverse,poco2018extracting,cliche2017scatteract,harper2014deconstructing}, and 
(iii) find the links between the text and parts of the chart~\cite{kong2014extracting, kim2018facilitating}. The latter is the least addressed, and the existing techniques are limited and do not automatically solve this task.

\figMotivation

In this work, we propose ChartText, a new method that automatically connects text with its respective area associated with the chart. Our method is the starting point for building a tool to help content creators generate interactive visualizations from static documents. Moreover, it is essential to clarify that we take for granted tasks (i) and (ii) to achieve this research result.
This approach can also work in semi-automatic mode, allowing the user to fix intermediate errors. Our method comprises two stages: In the matching stage, we generate individual links using different comparison operators between phrases in the text and textual elements in the chart. We combine the individual links in the grouping stage using each link's visual channel information. The grouped links, in general, are more meaningful and allow the creation of multiple applications. Here, we present two prototypes: In the first one, we make interactive documents using graphic overlays, facilitating the reading experience. We use voice instead of text to annotate charts in real-time in the second. For example, in a videoconference, our technique can automatically annotate a chart following the presenter's description.
We evaluate our results by measuring how similar the inferred links are to links manually annotated by experts in two datasets from academic and web documents. Also, we compare our results with KongExtraction~\cite{kong2014extracting}, a  non-automated method based on crowdsourcing, which works only with bar charts.
In summary, the contributions of this work are: 
\begin{itemize}
\item A method for linking text with visualization for multiple chart types. This method can work in two modes, fully automated and semi-automatic. 
\item An application to make interactive documents by visualizing the links generated by our technique. 
\item A proof of concept where we use audio instead of text to create video annotated charts in real-time.
\item A new dataset (text and charts) fully annotated for future comparisons. This dataset includes three chart types (bar chart, line chart, and scatter plots).
\end{itemize}

%% file: 2-related.tex
\section{Related Work}

Our work relates to two main areas: (i) chart interpretation and (ii) techniques and applications to facilitate the reading of documents.

\subsection{Chart Interpretation}
In recent years, methods and techniques have been developed to automatically extract the underlying data and visual encoding that build charts.
These techniques have applications such as the reuse and redesign of charts. One of the pioneers in this area is ReVision~\cite{savva2011revision}. 
ReVision extracts the data that generates a visualization from an image in a bitmap format.
ReVision proposed a pipeline where initially the images are classified using the Support Vector Machine model~\cite{SVMHearst1998} to determine the type of chart (\ie, bar chart, line chart, and scatter plot).
Then, they used different image processing techniques and heuristics. These methods are adapted to the chart type and then used to filter elements in the image and identify the data.
Other works maintain the initial idea of ReVision but vary some components with more sophisticated models. For instance, FigureSeer~\cite{siegel2016figureseer} and Scatterract~\cite{cliche2017scatteract} use a Convolutional Neural Network~\cite{cnnlecun}. 
The input of these systems consists of images and features based on the spatial information of the chart elements to enhance the classification task. 
They also take advantage of the axis information and text labels to infer the data on line charts and scatter plots. 
Later, ChartSense~\cite{jung2017chartsense} extended the data extraction support for five chart types (line chart, bar chart, area chart, foot chart, and radar chart), diversifying the heuristics to extract the data for each chart type but keeping the main idea proposed in the previous works.

Other works focused on recovering the visual encoding of the chart. 
For instance, Harper et al.~\cite{harper2014deconstructing} developed a system to deconstruct visualizations created with the D3~library~\cite{bostock2011d3}.
This technique uses the data and graphic elements, which are easier to access in SVG images, to compose the visual encoding. In the same vein, Poco and Heer~\cite{poco2017reverse} proposed a pipeline to recover the visual encoding but for bitmap images.
The technique exploits only the textual elements and retrieves a visual encoding in a declarative language similar to Vega-Lite~\cite{satyanarayan2017vega}. 
Later, Chen et al.~\cite{chen2019towards} presented a method to deconstruct bitmap infographics and generate templates that can be reused to create new infographics. 
This method uses deep learning models to identify graphics elements. 
Although we do not deconstruct charts in our work, we assume visual encoding as input. This assumption is valid since we can create visual encoding using the methods described above.

\subsection{Facilitating Reading of Documents}
Recently many works have appeared in this line. 
Kong and Agrawala~\cite{kong2012graphical} introduced Graphical Overlays to create interactive charts by accessing the data and visual encoding (using ReVision~\cite{savva2011revision}). 
We also have a similar application; however, we generate overlays based on the text accompanying a chart; in this sense, our goal is to facilitate document reading.

Other works take advantage of additional information that is accessible to enrich documents and visualizations. For instance, Metoyeret et al.~\cite{coupling-story} presented a technique that works on data-rich stories (\ie, texts with access to additional data such as tables). They proposed using text analysis to connect narrative sentences with visual elements that describe the rich data. Bryan et al.~\cite{temporal-summ} proposed a method to explore temporal data and find, through heuristics, points of interest in the data that allow the creation of explanatory visualizations. Hullman et al.~\cite{Contextifier} proposed a system that takes time series data on a company's stock to automatically create a line chart enriched with annotations from text analysis and metadata from news articles related to the company. These approaches are helpful when we have access to the data that accompanies the document or visualization. However, we only need the chart texts available in the visual encoding, and it is easy to extract with fewer errors in our approach.

Later, Kong et al.~\cite{kong2014extracting} presented a crowdsourcing-based method for extracting links between charts and text in documents. 
Nevertheless, they use people to annotate the document using crowdsourcing, unlike our work, which is fully automatic.
On the other hand, some works use tables instead of using visualizations. 
Kim et al.~\cite{kim2018facilitating} presented an automatic technique to find links between tables and paragraphs. 
This work uses NLP and heuristic techniques to connect table cells with text sentences. 
Similarly, Badam et al.~\cite{Badam2018ElasticDC} proposed a web tool that relates the text to the tables and creates visualizations using these links. 
To reduce the effort to create interactive documents, Latif et al.~\cite{linkingfra} proposed a framework for systematically generating documents that support text-visualization interaction. 
This system allows users to annotate the links in declarative language. 
Then, Cui et al.~\cite{Cui2019TexttoVizAG} proposed a method that converts text into infographics to facilitate reading. First, they identify parts of the text that refer to some data (\eg, numbers) and then use this information in templates that generate the image.
Later, Latif et al.~\cite{s.20191180} presented an approach for producing interactive documents for graph data, also using a declarative syntax that requires little to no programming skills. 
Recently, Lai et al.~\cite{autoannotation} proposed a system with similar capabilities. 
Nevertheless, they focus only on individual links; in our case, we have a second stage grouping the individual link to make more meaningful associations
Additionally, they require access to the encoded data, whereas our technique only requires the textual elements, representing a more straightforward task and less prone to error.

%% file: 3-data.tex
\section{Dataset}
\label{sec:dataset}
In this section, we first define the terminology used throughout the document. Then, we describe the data set supported by our technique and our assumptions. Finally, we depict how we create our dataset, the gold standard set in our evaluations.
\figVisualEncoding

\subsection{Definitions}
\label{section:definitions}
Before describing our technique, we must define some terms we often use throughout the document.

\myparagraph{Visual Encoding:}
  It describes how we create the visualization, i.e., converting data into a visual representation. 
  In a visual encoding, we specify the graphical marks (\eg, points, bar, line) and map data variables to the visual channels (\eg, position, length, size). 
  For instance,~\figref{fig:link_types}~(a) is an example of visual encoding that describes the channel associated with the x-axis of the line chart shown in~\figref{fig:link_types}~(b). 
  Note that we have the scale information in this specification\,---\,which maps values from the data domain to the image range. 
  Our visual encoding also has information about the labels used in the chart (text and bounding boxes) for this particular case. 

  Our technique assumes that we receive the visual encoding specification containing visual channel information (\ie, x-axis, y-axis, and color-legend), which can be automatically generated by methods proposed in previous works~\cite{poco2017reverse,harper2014deconstructing}.

\myparagraph{Textual Element:} It is a region on the chart that contains the text. For instance, in~\figref{fig:link_types}~(b), the x-axis has textual elements such as \phrase{100 million}, and the color legend has the textual element \phrase{deptype}.

\myparagraph{Phrase:} A phrase is any text fragment in a paragraph. For example, in~\figref{fig:link_types}~(b), the colored texts represent three phrases.

\myparagraph{Individual Link:} We define as an individual link a connection between a phrase and a textual element.
\figref{fig:link_types}~(b) shows some examples of individual links, each formed by a textual element and a phrase with the same number and color.

\myparagraph{Grouped Link:} A grouped link is a set of individual links. ChartText combines the links shown in~\figref{fig:link_types}~(b) to create a grouped link. Using this meaningful link, we can create a graphical overlay on the chart (more details in~\secref{section:grouping}).

\myparagraph{Case:} A case is a pair consisting of a chart and its associated paragraph. A case could contain many grouped links.

\subsection{Supported Visualizations}
Our method supports bar, line, and scatter plots as a result of the restrictions we defined to delimit the scope of this investigation:
i) the charts must represent the data using the cartesian space and a single color channel. We do not support visual channels such as orientation or size (\eg, it does not work for bubble charts); 
ii) we do not support charts composed of multiple views. Trellis plot is an example of a chart type violating this constraint; and
iii) we only consider charts that use a single visual mark. For instance, Pareto plots violate this rule because it uses two graphical marks: bars and lines.

\subsection{Gold Standard (GS) Set}
To evaluate our results, we compare our outcomes with the grouped links manually annotated by visualization experts\,---\,from now on, we will call it the Gold Standard (GS) set. Our cases come from two data sources: news websites and academic documents.

\myparagraph{Web Documents (\web):}
These are 24 cases taken from news websites and specialized platforms such as Pew Research (\url{pewresearch.org})  and The Economist (\url{economist.com}). 
This dataset was collected and annotated in Kong et al.~\cite{kong2014extracting} and is available on the paper's website.
The authors' complete datasets contain 40 cases, of which we chose 24 that meet our restrictions (all images are bar charts with their associated texts). 
Since our method requires visual encoding, we had to create the visual encoding manually.
%

\myparagraph{Academic Documents (\aca):}
This dataset has 75 cases extracted from the SemanticScholar (\url{semanticscholar.org/}) database, including figures and texts from scientific papers from computer science conferences such as ACM CHI and ACM UIST. 
The cases in this dataset include the three chart types: 25 bar charts, 25 line charts, and 25 scatter plots. 
Since this data had no annotations, we had to annotate the grouped links and visual encoding manually.
For this task, we follow the guideline proposed in Kong et al.~\cite{kong2014extracting}. In each paragraph-chart, we link the phrases with the textual elements. Each phrase must contain the minimum number of characters to refer to a textual element. For instance, in~\figref{fig:link_types}~(b), it would be possible to connect the phrase \phrase{DepType system} with the textual element \phrase{DepType} in the chart since adding the word \phrase{system} does not modify the meaning of the link. However, according to our guideline, we use \phrase{DepType} because it minimizes the selected text amount. The annotated groups comprise individual links (\ie, phrases and textual elements) that cover a fact entirely relating to text and chart. For instance, in~\figref{fig:link_types}~(b), the following fact is described in the text: \phrase{In the end with 100 million tokens, the accuracy of the Deptype system rises to 27\%.} and the individual links are highlighted in the paragraph and the chart (each individual link has the same color and numbering). If we remove any individual link in the example, the resulting group could not refer to the text's fact and would be meaningless. 
The two authors of this paper (experts in the visualization area) created the gold standard set. Each worked separately to generate the individual and grouped links using printed text and charts. Afterward,  they discussed the annotations, and the cases where they did not match were analyzed jointly (for this dataset, 30\% of the cases were discussed).

%% file: 4-technique.tex
\figTree
\section{Automatic Links Extraction}
We divide our method into two stages: matching and grouping. 
In the \emph{matching} stage, we make a series of comparisons between the paragraph text and the textual elements to find individual links. 
Then, in the \emph{grouping} stage, we combine the individual links that describe the same fact. 
We take advantage of the information in the individual links' visual channels in this stage.

\subsection{Matching}

This stage aims to link single phrases to textual elements to obtain individual links.
First, we split the paragraph into syntactically consistent sentences. (\eg, nominal phrases, verbal phrases, and prepositional phrases). 
For that, we decompose the paragraph into sentences using the sentence segmentation module of the spaCy library~\cite{spacy2}. 
Then, we compute the syntactic constituency tree for each sentence using the constituency parser method proposed by Kitaev et al.~\cite{Kitaev-2018-SelfAttentive}\,---\,the constituency tree subdivides the sentence into syntactic phrases. 
Each node in the tree represents a syntactic phrase; for instance,~\figref{fig:tree} shows the tree for the sentence \phrase{With 100 million tokens, the accuracy of the DepType system rises to 27.0\%}.
In this example, the first node with the label \emph{S} represents the first syntactic phrase corresponding to the complete sentence.
This phrase is then subdivided into three children nodes: \emph{PP}, \emph{NP} and \emph{VP} that correspond to three syntactic phrases, \emph{prepositional phrase}, \emph{noun phrase} and \emph{verbal phrase}, respectively.
After that, we traverse the tree and check whether the phrase associated with a node is related to any textual element in the chart. 
For instance, in~\figref{fig:tree}, the node \emph{QP}  (green box) corresponds to the phrase \phrase{100 million}, and it is related to the textual element \phrase{100 million} in the x-axis label. 
The same happens with the other two phrases in~\figref{fig:tree} (orange, purple and red boxes). 
Finally, we use different operators to compare the phrases with the textual elements: direct, unique word, semantic, and numerical value comparison.

\myparagraph{Direct Comparison.}
For each previously generated sentence, we check if any textual element is written precisely the same. If there is, then we create an individual link. This comparison helps identify proper nouns that are not defined in a vocabulary \eg, in~\figref{fig:tree}, the phrase \phrase{DepType} is the same as the textual element in the color legend.

\figUniqueWord

\myparagraph{Unique Word Comparison.}
It is common to have repeated terms in the textual elements with the same text role in the chart (see, for instance, legend labels in~\figref{fig:uniqueword}~(a).
This problem causes our direct comparison to generate false negatives because the phrases tend to reference the legend's terms that differentiate them (we call these terms unique words).
To solve this issue, we eliminate repeated words and keep the unique words on each textual element with the same text role.
In our example, we keep the terms \phrase{correct} and \phrase{incorrect} in the textual elements. These words are better candidates to match the phrases in the paragraph.

\myparagraph{Semantic Comparison.}
Sometimes phrases and textual elements are written using different words with the same meaning. 
For instance, in~\figref{fig:uniqueword}~(b), the phrase \phrase{religious wear} is related to the textual element \phrase{religious items or clothing}.
We compare the phrases and textual elements to cover these cases using a word embedding approach that captures the sentence's semantic. 
We tested multiple state-of-the-art techniques to convert phrases into high dimensional vectors; in the end, we opt for the Universal Sentence Encoder~\cite{cer-etal-2018-universal} technique because it got the best results in our datasets. We perform an analogous process for textual elements. 
Finally, we compute the similarity between a pair of vectors; we use the metric proposed by Cer et al.~\cite{cer-etal-2018-universal}, which calculates the cosine similarity and then use the $\arccos$ to convert the similarity into an angular distance ($\mathrm{sim}(\vect{u},\vect{v})=(1-\arccos(\frac{\vect{u}\cdot \vect{v}}{\|\vect{u}\|\|\vect{v}\|})/\pi)$). 
Experimentally they noticed that this technique performs better than basic cosine similarity.
We form an individual link if the similarity is more significant than a threshold of $0.78$\,---\,we set up this value by calculating the F1 score (described in~\secref{sec:eval}) on the web document data (WEB) using different thresholds. We use the threshold that returns the best score.

\figBadNumeric

\myparagraph{Numerical Value Comparison.}
In this comparison, we first identify the numerical values in the paragraph using SpaCy's Named Entity Recognition~\cite{spacy2} module\,---\,we only use the entities of type PERCENT, MONEY, CARDINAL, and DATE. We then verify the numerical values in the scales' domain in the chart (x/y axes). For this, we use channels with numerical variables in the visual encoding. If the numerical value is within any scale, we create an individual link.

The above procedure can generate multiple false positives presented as two types of errors. \figref{fig:badnumeric} shows these error types.

\myparagraph{Error type I.}
In~\figref{fig:badnumeric}~(a), the phrase \phrase{72\%} is within the y-axis domain. However, the \phrase{72\%} does not refer to the \phrase{accuracy} (y-axis) or any other axis.
To fix the first type error, we proceed as follows. First, we make sure that the numerical phrase has a NUM tag. Then, using the dependency tree, we find the NOUN associated with this number. After that, we verify if it coincides with some of the four cases shown in~\figref{fig:refnum1}~(a). Finally, once we have the NOUN phrase, we compare it with the titles of the chart's axes (we use the same comparisons of the previous stage: direct, unique, and semantic) and create the individual link; otherwise, we discard it. The four cases we show in~\figref{fig:refnum1}~(a) results from an extensive analysis with different documents to cover as many scenarios as possible. 

\myparagraph{Error type II.}
In~\figref{fig:badnumeric}~(b), we have an ``ambiguity" error, where the phrase \phrase{30\%} could belong to both the x-axis and the y-axis. However, reading the paragraph deduces that it refers to the x-axis (\phrase{Coverage PC}). To solve these problems, we generate a data structure that contains two types of information for each word: i) the syntactic dependency tree and ii) the Part of Speech (POS) tags.
To fix the second type of error, we proceed as follows. Remember that the numerical phrase is associated with both axes (ambiguity). Then, we look for the axes' titles (textual elements) and the numerical phrase in the tree. This search uses the same comparison as the previous steps. In~\figref{fig:refnum1}~(b), we can see an example for~\figref{fig:badnumeric}~(b); text in red is the numerical phrase, and the titles of the x-axis and y-axis are in green and blue, respectively. After that, we calculate the paths between the numerical phrase and the axes titles in the dependency tree (in~\figref{fig:refnum1}~(b), we can see these paths with blue and green lines). Finally, we select the phrase that has the shortest route.  In our example, we will choose the phrase \phrase{Coverage PC}, discarding the link with the y-axis.

\figRefinement

\subsection{Grouping}
\label{section:grouping}
While individual links can already help generate interactive documents, as presented in Lai et al.~\cite{autoannotation}. Combining individual links can get more meaningful relationships related to a narrative aspect.
For instance,~\figref{fig:link_types}~(b) shows the individual links; with that, we can only highlight the areas where the textual elements and phrases are located in the chart or paragraph, respectively. 
In this example, phrases \phrase{100 million},  \phrase{27.0\%}, and \phrase{DepType} correspond to the \texttt{x-position}, \texttt{y-position}, and \texttt{color} channels, respectively. However, if we combine them, as the phrase \phrase{In the end, with 100 million tokens, the accuracy of the DepType system rises to 27.0\%} suggests, we can find a specific location in the plotting area.  
To do so, we noticed that in our Gold Standard set, the grouped links annotated by the experts usually combine textual elements from different visual channels (\texttt{x-position}, \texttt{y-position}, and \texttt{color}). Thus, our method for grouping attempts to imitate how a human would do it.

\figGroups

\myparagraph{Transferring of the Visual Channel to the Phrases.} 
The first step is to transfer the visual channel information (\ie, \texttt{x-position}, \texttt{y-position}, or \texttt{color}) from the textual elements to the associated phrases in the links. 
For example, in~\figref{fig:tree}, the phrase \phrase{100 million} is linked to the textual element corresponding to an x-axis label. 
Therefore, we transfer the channel x-position to that phrase (\phrase{100 million}). We apply the same procedure for the other sentences (\phrase{Deptype} and \phrase{27.0}).

\myparagraph{Searching Groups in Constituency Tree.}
We group the individual links whose phrases are close enough in the constituency tree and are associated with different visual channels.
To achieve this, nodes associated with individual link phrases inherit the phrase's visual channel (see~\figref{fig:groups} as an example). 
To form the groups, we traverse the constituency tree, and for each node, we count the number of visual channels associated with its descendants. 
Then, we choose those nodes that maximize the variability of visual channels, minimize the number of children nodes, and maximize the tree level\,---\,this last condition makes the sentences shorter. 
In~\figref{fig:groups}, the highlighted region shows how the node S is selected to form a group (group 2). The node \emph{S}  meets the conditions to form a group because it is an ancestor of nodes \emph{NP}, \emph{ADJP}, and \emph{NP}  associated with visual channels \texttt{x-position}, \texttt{color}, and \texttt{y-position}. Finally, we combines the individual links related to this node and create a \emph{grouped link}.

\figSimilitud

\figResult

%% file: 5-evaluation.tex
\section{Evaluation}
In this section, we first describe the metric used for our evaluations. Then, we discuss some quantitative and qualitative results to reason about good and bad scenarios.

\subsection{Metric}\label{sec:eval}
Our work's metric adapts the metrics proposed in object detection in images and information retrieval in texts. The objective is to evaluate how similar our grouped links are to the Gold Standard (GS) set. \figref{fig:similitud} will serve as a reference to explain each stage.
The F1 score gives us a value of 1 when two sets are identical and 0 when they are different. To interpret this score, let us see~\figref{fig:erroneous} as an example. On the right side of the figure, we can see the links on the GS, and on the left side, the links obtained by our method. If we had found all the correct links, we would have an F1 score of 1, but we failed to find the textual element for the phrase \phrase{84\%}\,---\,ChartText wrongly assigned the phrase with a numeric value in the x-axis, when in fact it belongs to the y-axis\,---\,causing the F1 score to decrease to 88\%. As the errors increase, the score decrease, \eg, if we remove the links related to the phrases \phrase{10\%} and \phrase{active agents} we will obtain a score of 66\%.
As we can see, each error in an individual link is, on average, a 12\% error, which gives us a clearer idea of how our method is behaving in all the cases.

To facilitate the description, we concentrate on a single \emph{case}, \ie, a set of grouped links. In \figref{fig:similitud}(a), the blue and red circles represent the grouped links of ChartText and the GS set, respectively. Ideally, there should be a one-to-one pairing; however, this seldom happens. The similarity value we use is the $F1= \frac{2 \cdot precision\cdot recall}{precision + recall}$ score; therefore, we need to calculate the precision and recall. 
To calculate the precision, we take each grouped link from ChartText and look for the most similar grouped link in the GS set\,---\,later, we will explain what we mean by the most similar. Once we have the matching, we add their similarities and divide them by the total number of pairs (for our example in (a), the precision is 0.65). Intuitively, the precision tells us how good are the grouped links generated by ChartText. 
Similar reasoning applies for recall but goes from the GS set to the grouped links generated by ChartText. (in our example in (b), the recall is 0.47). Intuitively, this value tells us to recover all the grouped links from the GS set. Finally, the F1 score for our case is 0.55.

In the previous paragraph, we mentioned that we look for the most similar grouped link in the other set given a grouped link. To assign a similarity value between grouped links, we first calculate the similarity between textual elements and phrases. The average of these values will be the similarity of the grouped link. Next, we explain how to compute both values. 

\myparagraph{Textual Elements Similarity.}
Here, we also use the F1 score, and the procedure is similar to before but using the textual elements from the ChartText and GS set.  For precision, we take each textual element from ChartText and look for its corresponding one in the GS set (this value will be 1 or 0, if it is correct or not), then we add these values and divide them by the number of textual elements in ChartText. For our example in (c), the precision is 1.  Again, similar reasoning applies to the recall. For our example in (c), the recall is 0.5, and the F1 score is 0.66.

\myparagraph{Phrases Similarity.}
Similarly, we use the F1, precision, and recall scores as in the previous paragraph. The difference is that we use two sets of phrases, and the similarity between phrases is calculated with the Intersection over Union (IoU) metric since the sentences generated by ChartText do not precisely match the GS set. In our example in (d), we have a precision of 0.4, a recall of 0.5, and an F1 score of 0.5.

\subsection{Quantitative Results}
\figPhasesEval
Using the metrics described above, we performed a series of experiments to validate the results generated by ChartText.

\myparagraph{Comparison with KongExtraction~\cite{kong2014extracting}.}
We performed this comparison with the \web dataset, which is the same used by KongExtraction. Our method obtains an average similarity of 50\%, while KongExtraction obtains 58\%.~\figref{fig:result_1}~(a) shows the similarity for each case in this dataset. It is important to note that although KongExtraction has a higher similarity, ChartText has certain benefits: i) our approach is fully automatic, while KongExtraction requires human participation through crowdsourcing; ii) KongExtraction needs to access the underlying data of the chart, while ChartText only requires access to the text present in the chart (available in the visual encoding), which is a simpler and less error-prone task in automation.
Additionally, we evaluate ChartText with the \aca dataset, and we get an average similarity of 66\%.~\figref{fig:result_1}~(b) shows the similarity for each case in this dataset. In this experiment, we could not compare with KongExtraction since its code is unavailable, and it would be necessary to recreate the crowdsourcing environment. 

\myparagraph{Evaluating phases for individual links.}
This experiment's objective is to validate the usefulness of each of the comparisons in the matching stage. For that, we compute our metrics after each comparison (\ie, before grouping): Direct, Unique Word, Semantic, and Numerical-value. Given that we do not have groups at this stage yet, we need to distribute the grouped links of the GS set into individual links. In~\figref{fig:phaseseval}, we can see the results of this experiment. As we advance in the pipeline, the F1, precision, and recall scores increase, validating each comparison's need. However, the precision score decreases slightly in the last comparison (Numerical-value). To explain this, let's remember that our method finds new individual links in each step, and some of them can be false positives. Nevertheless, this effect is compensated by an increase in the recall since we bring individual links with numerical values necessary for the grouping stage and the applications. 

\myparagraph{Semi-automatic evaluation.}
On the other hand, it is possible to use ChartText in a semi-automatic mode where the user can intervene to solve specific problems (more details about human intervention in the Applications section). 
In the semi-automatic mode, the user can manipulate and fix individual links so that the grouping stage will get better results. 
We calculate each case's similarities to validate this statement, assuming that we have the perfect individual links. 
Our results show similarities increased from 50\% to 68\% for the \web dataset and from 66\% to 84\% for the \aca dataset.

\figErroneous
\subsection{Qualitative Results}
In this section, we present qualitative observations to explain the behavior in some relevant cases, taking into account the design of our method.

\myparagraph{Well-Performed Results.}
ChartText has good results when it is easy to associate the phrases with the visual channels (x-position, y-position, and color). 
For instance, In \figref{fig:overlay_example}, the phrase \phrase{Oracle} is easily associated with a color legend label (color channel) in the chart. 
Similarly, the phrase \phrase{96.23} is directly associated with a value on the y-axis (y-position channel), without any chance of being wrong because the x-axis only admits values less than or equal to 10. 
Finally, the phrase \phrase{k = 10} explicitly indicates that the value 10 belongs to the axis with the title “k” which is the x-axis (x-position channel). 
In this example, we notice that the three phrases that form the grouped link belongs to the short phrase \phrase{the Oracle achieve 96.23\% when k = 10}. 
We designed our method to create grouped links with multiple channels, and the phrases are close to each other.
This behavior emulates how experts annotated the grouped links following this same criterion.

\myparagraph{Poorly Performed Results.}
ChartText does not get correct results when there is ambiguity in assigning the visualization role to a phrase. 
Consider the case illustrated in \figref{fig:erroneous}~(a); On the left side, we can see the wrong output generated by ChartText, and on the right side, we can see the correct overlays generated with the grouped links in the Gold Standard Set. 
Analyzing this case, notice that ChartText identified all the phrases correctly; however, the phrase \phrase{84} was incorrectly associated with the x-axis.
This error happens in the matching stages. 
The phrase \phrase{84} is assigned to the x/y-axis because the value 84 fit in both scale's domains. 
Then, using our rule-based approach and the syntactic distances to the phrases, ChartText determines that the phrase \phrase{84} belongs to the axis associated with the nearby phrase \phrase{active agents} (x-axis).
In other words, our approach could not identify the attribute (NOUN) related to the numerical phrase. 
For this particular example, we can see in the text that no phrase indicates that the \phrase{84\%} refers to the y-axis (\phrase{Cumulative probability}).

Another problematic case is shown in \figref{fig:erroneous}~(b). 
The main problem is that ChartText requires additional context information to reference a phrase with a textual element. 
In the GS set, the experts found two phrases associated with the chart. 
For instance, experts decided that the phrase \phrase{40-something} means people between 40 and 49. It makes sense to link this phrase with the textual element \phrase{35-49} since it includes this age range.
The second phrase, \phrase{retires}, was associated with people over 65, which appears in the chart as \phrase{65+}. ChartText would need additional information to generate this link, such as the meaning of the phrase \phrase{40-something}. It is a challenging problem and requires more sophisticated models. 


%% file: 6-apps.tex
\section{Applications}

To demonstrate the usefulness of ChartText, we developed two applications: (1) A interactive document web tool and (2) Voice2Overlays, a tool that turns voice into overlays.

\subsection{Interactive Documents}
Using CharText, we build a proof-of-concept that generates interactive overlays from a chart in bitmap format based on a text that describes it. 
This application exemplifies how our method would allow us to build tools that generate interactive documents. 
This idea can be easily extrapolated in other more valuable applications in the real world. 
For instance, a document editor allows a content creator (such as a web news writer) to embed a chart on a text and generate interactive documents with minimal effort and without programming knowledge. 
Even at a more practical level, a plugin for a browser or a document reader (\eg, Adobe reader~\cite{adobeacrobat}) that allows the end-users to select the graphic and the text on which they wish to inquire and automatically generate the highlights between the chart and the phrases of the corresponding text.

\myparagraph{Overlays Generation.}
We built a web tool that supports text visualization interaction to show the overlays.
\figref{fig:overlay_example} shows an example of how the tool produces an overlay when the user selects a phrase in the text. 
In the example, the tool highlights a point at the end of the curve that corresponds to the color legend \phrase{Oracle} (intersection of red lines).
This overlay corresponds to three phrases (highlighted in yellow): \phrase{Oracle}, \phrase{K=10}, and \phrase{96.23\%}. 
These phrases belong to individual links in the same group, and we use the visual channel associated with each link to build the overlay.
For instance, we use the color information associated with the phrase \phrase{Oracle} to filter out only the pixels that correspond to that color in the chart. 
Using this strategy, we highlight the correct curve in the chart.
To identify the point at the end of the curve, we use the spatial information associated with the phrases \phrase{K=10} and \phrase{96.23}. 
Different overlays can be generated, depending on the type of chart and the visual channel associated with the group's links.
Now we explain the six overlay types supported by our application: 
\figOverlayApp

\figOverlays

\begin{description}[style=unboxed,leftmargin=0cm]
  \item[a) Line Segment.]
This overlay is used to represent a single value of an axis. We use the numerical value in the grouped link to draw a line segment along the chart's width or height. The orientation depends on the visual channel associated with the textual element (x-position or y-position). 
\figref{fig:lineoverlays}~(a), shows an example when the numerical value belongs to the x-axis.

  \item[b) Target Point.]
This type of overlay is generated when, in the grouped link, there are two textual elements related to two different chart numerical axes. In this case, as shown in \figref{fig:lineoverlays}~(b), we draw two line segments from the axes (orthogonal directions). The intersection represents the target point to which the phrase refers.

  \item[c) Color Filter.]
When a phrase in the grouped link refers to a color legend on the chart, we filter out the chart's pixels with other colors, highlighting only the relevant color. To achieve this, we increase the transparency of the pixels that do not correspond to the target color (\figref{fig:lineoverlays}~(c)).

  \item[d) Bounding Box.]
We use this type of overlay to highlight a bar or a group of bars, as shown in \figref{fig:lineoverlays}~(d). The Bounding Box is used when only one individual link refers to a  nominal axis (only for bar charts).

  \item[e) Highlighted Bar.]
This type of overlay is proper when we want to highlight a bar (\figref{fig:lineoverlays}~(e)). 
We generate this overlay when there are two phrases in the grouped link: one related to a color legend and another related to a numerical axis label. 
To create this overlay, we increase the pixels' transparency that does not correspond to the target color (we could still have multiple bars with the same color). 
Then, we increase the pixels' transparency that does not correspond to the axis label in the grouped link, reducing it to one highlighted bar.

  \item[f) Target Bar.]
This overlay is a specialization of the target point but applied to bar charts.
\figref{fig:lineoverlays}~(f) shows an example of this type of overlay. 
It is generated when the grouped link contains textual elements from all three visual channels (x-position, y-position, and color). We need to highlight the bar and the numerical value associated with the bar. 
We first highlight the bar using the same strategy to generate the Highlighted Bar to create this overlay. 
Then, we draw a line segment representing the numerical value of the axis.
\end{description}

\myparagraph{Semi-automatic Tool.}
Our tool has a semi-automatic mode that allows users to interact with the interface to fix intermediate results.
Specifically, the user can verify, correct, or remove individual links. Once the user is satisfied with the results, ChartText continues with the grouping stage and generates the overlays. 

For example, in~\figref{fig:semiauto}~(a), we correct the problem presented in~\figref{fig:erroneous}~(a). When we double click on the phrase \phrase{84\%}, the system displays a panel where it is possible to select the textual element associated with the phrase.
In this particular case, ChartText erroneously linked the phrase \phrase{84\%} with an x-axis; however, we fix this link, choosing the correct value in the \texttt{y-axis}.
By making this change and executing the grouping stage, ChartText manages to generate the correct grouped link, which is displayed as an overlay in~\figref{fig:erroneous}~(a-right)
Similarly, we can fix the error presented in~\figref{fig:erroneous}~(b); we double click the \phrase{retires} in the text to pop-up the panel in~\figref{fig:semiauto}~(b), then we select the correct textual element (\phrase{+65}).

\figSemiauto

\subsection{Voice2Overlays}
We can use ChartText as the primary system component to convert voice to overlays. This functionality provides interactivity to static charts in environments such as conferences and video calls. 
The application receives as input the audio containing the reference to the chart. Then, we use a speech-to-text tool to convert audio into text. 
We take advantage of these tools to detect pauses in speech and generate sentences;
This segmentation is beneficial since our method works better with short phrases, improving the analysis during the grouping stage.
This feature also helps us determine how long an overlay should remain in view. For this, we take the time when the API generates a new sentence as a reference. We add the new overlay and remove the previous one based on it.
Once we have the sentence, we use ChartText to get the grouped links and generate the graphical overlays.

\figVoiceOverlaysOne

To demonstrate the usefulness of this application, we decided to conduct a proof of concept. 
We selected two videos from Ted Talks conferences, where presenters explain a chart to the audience, an ideal scenario for our application. 
We extracted the audio fragment for each video and reconstructed the chart's visual encoding shown in the video. 
Then, we use the Google~speech-to-text~API~\cite{google_cloud_speech_to_text} to convert the audio to text. Finally, we use our method to extract the links and generate the overlays for each sentence in the audio.

The first video is entitled \phrase{Why you should love statistics}; in the video, the presenter uses a bar chart to explain the percentage of young people with low numeracy in 12 countries. \figref{fig:app2} shows the results obtained for the first video. \figref{fig:app2}-top shows the frames of the original video and the sentences referring to the chart. \figref{fig:app2}-bottom shows the overlays on the chart generated by ChartText. For instance, for the first phrase \phrase{Leading the way the USA nearly 40 percent}, there is an overlay that highlights the first bar of the chart. We can see the results of this example in the accompanying video. 

The second video is entitled \phrase{Germany: Low Crime, Clean Prisons, Lessons for America}. The presenter explains the U.S. incarceration rate between 1925 and 2012 using a scatter plot in this video. \figref{fig:v2o_case2} shows the results obtained. 
The top part shows the sentences extracted from the original video with a reference frame as in the first video. The bottom part shows the overlays generated by our method. 
In this example, we obtained three overlays: The first overlay highlights a date range (\phrase{1925 to 1975}) on the x-axis and a value (\phrase{100}) on the y-axis. 
The second and third overlays highlight specific dates on the x-axis (\phrase{1940} and \phrase{1970}, respectively).

%% file: 7-limitations.tex
\section{Limitations and Future Work}
Our work has some limitations and possible improvements presented in this section.

\myparagraph{Dataset.}
One of the significant limitations of this work is the small amount of annotated data available to relate the text to visualizations. 
In this work, we are making another fully annotated dataset, including three chart types; however, this is still insufficient to archive this task successfully. 
It is impossible to explore other techniques and approaches without an extensive dataset, such as using machine learning models instead of the rule-based system currently used.
As future work, we propose to design the processes of annotating an extensive dataset to relate text and visualizations.

\myparagraph{Improvement in the Numerical-value phase.}
As we saw in our evaluation, the Numerical-value comparison is necessary for the grouping stage, affecting the generation of overlays.
However, our method is not robust enough, generating false positives in the individual links. 
The main problem lies in using heuristics and templates to identify the numerical phrases that do not correspond to the chart's axes.
In future work, we intend to experiment with more robust methods from natural language processing. 
We will have to generate a larger dataset with annotated text examples. 
Using it, we could train a machine learning model to identify the NOUNs associated with the numerical phrases more precisely without rules.

\myparagraph{Support for complex charts}.
In this research, we limited the scope of ChartText to simple charts but were commonly used to display results with descriptive statistics (bar charts, line charts, and scatterplots). 
However, this is a starting point for future works that perform the same task on more sophisticated graphics (i.e., that overload graphic markup and visual channels). 
For instance, there are two graphic marks in a Pareto chart: bars and lines. Our technique does not support this chart type but supports each type separately. A possible extension of our approach, having the data properly annotated, would help solve this problem.

\figVoiceOverlaysTwo

\myparagraph{End-to-end Generation of Interactive Documents.}
ChartText assumes that we have: i) visual encoding and ii) chart and text location within a document. 
These are valid assumptions since previous works~\cite{pdffigures2, mapAnalysis2018,poco2018extracting,harper2014deconstructing} address these problems with relatively good results under certain conditions. 
In this work, we decided not to enter into those areas because we focused on finding the links between the text and the chart, which is already difficult.
As future work, we propose implementing and integrating all the components mentioned above to build a fully automatic system that takes as input a document (in PDF format) and generates an interactive document.
With this framework, it will be possible to implement plugins for web browsers or document readers tools such as Adobe Acrobat Reader~\cite{adobeacrobat}.

%% file: 8-conclusion.tex
\section{Conclusions}
This paper presents ChartText, a method automatically extracting links between text and visualizations. We evaluated our results by comparing them with manual annotations made by experts on two datasets (web and academic documents).
We compare our ChartText with KongExtraction~\cite{kong2014extracting} (a non-automatic method based on crowdsourcing strategy) for the web documents.
Our approach obtained a similarity of 50\% compared to 58\% of KongExtraction. Despite our lower score, it is essential to highlight that: i) our method provides a fully automatic approach, ii) we support more chart types, and iii) we do not need access to the underlying data; we only need the charts' textual elements. 
We also make a second dataset containing 75 cases, including three chart types from academic documents. In this dataset, Chartext got an average similarity of 66\%.
Although our method is automatic, we have a semi-automatic mode that allows the user to fix intermediate errors (individual links). We experimented with perfect individual links to evaluate this functionality, and our scores rose to 68\% and 84\% for the web and academic documents, respectively.
Finally, we presented two applications. The first is a tool to generate interactive documents. The second is Voice2Overlays, which converts a speaker's voice into visual aids on a chart.

%% file: paper.bbl
\begin{thebibliography}{10}

\bibitem{adobeacrobat}
Adobe.
\newblock Adobe acrobat reader.
\newblock \url{https://get.adobe.com/}, 2019.
\newblock Accessed: 2019-10-04.

\bibitem{Badam2018ElasticDC}
S.~K. Badam, Z.~Liu, and N.~Elmqvist.
\newblock Elastic documents: Coupling text and tables through contextual
  visualizations for enhanced document reading.
\newblock {\em IEEE Transactions on Visualization and Computer Graphics},
  25:661--671, 2018.

\bibitem{bostock2011d3}
M.~Bostock, V.~Ogievetsky, and J.~Heer.
\newblock D3 data-driven documents.
\newblock {\em IEEE Transactions on Visualization and Computer Graphics},
  17(12):2301--2309, Dec. 2011. doi: {{%
10\hspace{.1pt}\discretionary{.}{%
}{.}\hspace{.4pt}1109\discretionary{/}{%
}{/}TVCG\hspace{.1pt}\discretionary{.}{%
}{.}\hspace{.4pt}2011\hspace{.1pt}\discretionary{.}{%
}{.}\hspace{.4pt}185}}


\bibitem{temporal-summ}
C.~Bryan, K.-L. Ma, and J.~Woodring.
\newblock Temporal summary images: An approach to narrative visualization via
  interactive annotation generation and placement.
\newblock {\em IEEE Transactions on Visualization and Computer Graphics},
  23(1):511–520, Jan. 2017. doi: {{%
10\hspace{.1pt}\discretionary{.}{%
}{.}\hspace{.4pt}1109\discretionary{/}{%
}{/}TVCG\hspace{.1pt}\discretionary{.}{%
}{.}\hspace{.4pt}2016\hspace{.1pt}\discretionary{.}{%
}{.}\hspace{.4pt}2598876}}


\bibitem{cer-etal-2018-universal}
D.~Cer, Y.~Yang, S.-y. Kong, N.~Hua, N.~Limtiaco, R.~St.~John, N.~Constant,
  M.~Guajardo-Cespedes, S.~Yuan, C.~Tar, B.~Strope, and R.~Kurzweil.
\newblock Universal sentence encoder for {E}nglish.
\newblock In {\em Proceedings of the 2018 Conference on Empirical Methods in
  Natural Language Processing: System Demonstrations}, pp. 169--174.
  Association for Computational Linguistics, Brussels, Belgium, Nov. 2018. doi:
  {{%
10\hspace{.1pt}\discretionary{.}{%
}{.}\hspace{.4pt}18653\discretionary{/}{%
}{/}v1\discretionary{/}{%
}{/}D18\discretionary{%
}{-}{-}2029}}


\bibitem{chen2019towards}
Z.~Chen, Y.~Wang, Q.~Wang, Y.~Wang, and H.~Qu.
\newblock Towards automated infographic design: Deep learning-based
  auto-extraction of extensible timeline.
\newblock {\em IEEE transactions on visualization and computer graphics}, 2019.

\bibitem{pdffigures2}
C.~Clark and S.~Divvala.
\newblock Pdffigures 2.0: Mining figures from research papers.
\newblock In {\em Proceedings of the 16th ACM/IEEE-CS on Joint Conference on
  Digital Libraries}, JCDL '16, p. 143–152. Association for Computing
  Machinery, New York, NY, USA, 2016. doi: {{%
10\hspace{.1pt}\discretionary{.}{%
}{.}\hspace{.4pt}1145\discretionary{/}{%
}{/}2910896\hspace{.1pt}\discretionary{.}{%
}{.}\hspace{.4pt}2910904}}


\bibitem{cliche2017scatteract}
M.~Cliche, D.~Rosenberg, D.~Madeka, and C.~Yee.
\newblock Scatteract: Automated extraction of data from scatter plots.
\newblock In M.~Ceci, J.~Hollm{\'e}n, L.~Todorovski, C.~Vens, and
  S.~D{\v{z}}eroski, eds., {\em Machine Learning and Knowledge Discovery in
  Databases}, pp. 135--150. Springer International Publishing, Cham, 2017.

\bibitem{Cui2019TexttoVizAG}
W.~Cui, X.~Zhang, Y.~Wang, H.~Huang, B.~Chen, L.~Fang, H.~Zhang, J.-G. Lou, and
  D.~Zhang.
\newblock Text-to-viz: Automatic generation of infographics from
  proportion-related natural language statements.
\newblock {\em IEEE transactions on visualization and computer graphics}, 2019.

\bibitem{echeverria2009external}
M.~d. P.~P. Echeverr{\'\i}a and N.~Scheuer.
\newblock External representations as learning tools: An introduction.
\newblock In {\em Representational systems and practices as learning tools},
  pp. 1--18, 2009.

\bibitem{google_cloud_speech_to_text}
Google.
\newblock Cloud speech-to-text api.
\newblock \url{https://cloud.google.com/speech-to-text/docs/apis}, 2019.
\newblock Accessed: 2019-10-04.

\bibitem{harper2014deconstructing}
J.~Harper and M.~Agrawala.
\newblock Deconstructing and restyling d3 visualizations.
\newblock In {\em Proceedings of the 27th Annual ACM Symposium on User
  Interface Software and Technology}, UIST '14, pp. 253--262. ACM, New York,
  NY, USA, 2014. doi: {{%
10\hspace{.1pt}\discretionary{.}{%
}{.}\hspace{.4pt}1145\discretionary{/}{%
}{/}2642918\hspace{.1pt}\discretionary{.}{%
}{.}\hspace{.4pt}2647411}}


\bibitem{SVMHearst1998}
M.~A. Hearst.
\newblock Support vector machines.
\newblock {\em IEEE Intelligent Systems}, 13(4):18--28, July 1998. doi: {{%
10\hspace{.1pt}\discretionary{.}{%
}{.}\hspace{.4pt}1109\discretionary{/}{%
}{/}5254\hspace{.1pt}\discretionary{.}{%
}{.}\hspace{.4pt}708428}}


\bibitem{spacy2}
M.~Honnibal and I.~Montani.
\newblock spacy 2: Natural language understanding with bloom embeddings,
  convolutional neural networks and incremental parsing.
\newblock {\em To appear}, 2017.

\bibitem{Contextifier}
J.~Hullman, N.~Diakopoulos, and E.~Adar.
\newblock Contextifier: Automatic generation of annotated stock visualizations.
\newblock In {\em Proceedings of the SIGCHI Conference on Human Factors in
  Computing Systems}, CHI '13, p. 2707–2716. Association for Computing
  Machinery, New York, NY, USA, 2013. doi: {{%
10\hspace{.1pt}\discretionary{.}{%
}{.}\hspace{.4pt}1145\discretionary{/}{%
}{/}2470654\hspace{.1pt}\discretionary{.}{%
}{.}\hspace{.4pt}2481374}}


\bibitem{jung2017chartsense}
D.~Jung, W.~Kim, H.~Song, J.-i. Hwang, B.~Lee, B.~Kim, and J.~Seo.
\newblock Chartsense: Interactive data extraction from chart images.
\newblock In {\em Proceedings of the 2017 CHI Conference on Human Factors in
  Computing Systems}, CHI '17, pp. 6706--6717. ACM, New York, NY, USA, 2017.
  doi: {{%
10\hspace{.1pt}\discretionary{.}{%
}{.}\hspace{.4pt}1145\discretionary{/}{%
}{/}3025453\hspace{.1pt}\discretionary{.}{%
}{.}\hspace{.4pt}3025957}}


\bibitem{kim2018facilitating}
D.~H. Kim, E.~Hoque, J.~Kim, and M.~Agrawala.
\newblock Facilitating document reading by linking text and tables.
\newblock In {\em Proceedings of the 31st Annual ACM Symposium on User
  Interface Software and Technology}, UIST '18, pp. 423--434. ACM, New York,
  NY, USA, 2018. doi: {{%
10\hspace{.1pt}\discretionary{.}{%
}{.}\hspace{.4pt}1145\discretionary{/}{%
}{/}3242587\hspace{.1pt}\discretionary{.}{%
}{.}\hspace{.4pt}3242617}}


\bibitem{Kitaev-2018-SelfAttentive}
N.~Kitaev and D.~Klein.
\newblock Constituency parsing with a self-attentive encoder.
\newblock In {\em Proceedings of the 56th Annual Meeting of the Association for
  Computational Linguistics (Volume 1: Long Papers)}. Association for
  Computational Linguistics, Melbourne, Australia, July 2018.

\bibitem{kong2012graphical}
N.~Kong and M.~Agrawala.
\newblock Graphical overlays: Using layered elements to aid chart reading.
\newblock {\em IEEE Transactions on Visualization and Computer Graphics},
  18(12):2631--2638, Dec 2012. doi: {{%
10\hspace{.1pt}\discretionary{.}{%
}{.}\hspace{.4pt}1109\discretionary{/}{%
}{/}TVCG\hspace{.1pt}\discretionary{.}{%
}{.}\hspace{.4pt}2012\hspace{.1pt}\discretionary{.}{%
}{.}\hspace{.4pt}229}}


\bibitem{kong2014extracting}
N.~Kong, M.~A. Hearst, and M.~Agrawala.
\newblock Extracting references between text and charts via crowdsourcing.
\newblock In {\em Proceedings of the SIGCHI Conference on Human Factors in
  Computing Systems}, CHI '14, pp. 31--40. ACM, New York, NY, USA, 2014. doi:
  {{%
10\hspace{.1pt}\discretionary{.}{%
}{.}\hspace{.4pt}1145\discretionary{/}{%
}{/}2556288\hspace{.1pt}\discretionary{.}{%
}{.}\hspace{.4pt}2557241}}


\bibitem{autoannotation}
C.~Lai, Z.~Lin, R.~Jiang, Y.~Han, C.~Liu, and X.~Yuan.
\newblock Automatic annotation synchronizing with textual description for
  visualization.
\newblock In {\em Proceedings of the 2020 CHI Conference on Human Factors in
  Computing Systems}, CHI '20, p. 1–13. Association for Computing Machinery,
  New York, NY, USA, 2020. doi: {{%
10\hspace{.1pt}\discretionary{.}{%
}{.}\hspace{.4pt}1145\discretionary{/}{%
}{/}3313831\hspace{.1pt}\discretionary{.}{%
}{.}\hspace{.4pt}3376443}}


\bibitem{linkingfra}
S.~Latif, D.~Liu, and F.~Beck.
\newblock Exploring interactive linking between text and visualization.
\newblock 06 2018. doi: {{%
10\hspace{.1pt}\discretionary{.}{%
}{.}\hspace{.4pt}2312\discretionary{/}{%
}{/}eurovisshort\hspace{.1pt}\discretionary{.}{%
}{.}\hspace{.4pt}20181084}}


\bibitem{s.20191180}
S.~Latif, K.~Su, and F.~Beck.
\newblock {Authoring Combined Textual and Visual Descriptions of Graph Data}.
\newblock In J.~Johansson, F.~Sadlo, and G.~E. Marai, eds., {\em EuroVis 2019 -
  Short Papers}. The Eurographics Association, 2019. doi: {{%
10\hspace{.1pt}\discretionary{.}{%
}{.}\hspace{.4pt}2312\discretionary{/}{%
}{/}evs\hspace{.1pt}\discretionary{.}{%
}{.}\hspace{.4pt}20191180}}


\bibitem{cnnlecun}
Y.~LeCun, Y.~Bengio, and G.~Hinton.
\newblock Deep learning.
\newblock {\em Nature}, 521:436--44, 05 2015. doi: {{%
10\hspace{.1pt}\discretionary{.}{%
}{.}\hspace{.4pt}1038\discretionary{/}{%
}{/}nature14539}}


\bibitem{mapAnalysis2018}
A.~Mayhua, E.~Gomez-Nieto, J.~Heer, and J.~Poco.
\newblock Extracting visual encodings from map chart images with color-encoded
  scalar values.
\newblock {\em Conference on Graphics, Patterns and Images, 31. (SIBGRAPI)},
  2018.

\bibitem{coupling-story}
R.~Metoyer, Q.~Zhi, B.~Janczuk, and W.~Scheirer.
\newblock Coupling story to visualization: Using textual analysis as a bridge
  between data and interpretation.
\newblock In {\em 23rd International Conference on Intelligent User
  Interfaces}, IUI '18, p. 503–507. Association for Computing Machinery, New
  York, NY, USA, 2018. doi: {{%
10\hspace{.1pt}\discretionary{.}{%
}{.}\hspace{.4pt}1145\discretionary{/}{%
}{/}3172944\hspace{.1pt}\discretionary{.}{%
}{.}\hspace{.4pt}3173007}}


\bibitem{poco2017reverse}
J.~Poco and J.~Heer.
\newblock Reverse-engineering visualizations: Recovering visual encodings from
  chart images.
\newblock {\em Comput. Graph. Forum}, 36(3):353--363, June 2017. doi: {{%
10\hspace{.1pt}\discretionary{.}{%
}{.}\hspace{.4pt}1111\discretionary{/}{%
}{/}cgf\hspace{.1pt}\discretionary{.}{%
}{.}\hspace{.4pt}13193}}


\bibitem{poco2018extracting}
J.~Poco, A.~Mayhua, and J.~Heer.
\newblock Extracting and retargeting color mappings from bitmap images of
  visualizations.
\newblock {\em IEEE transactions on visualization and computer graphics},
  24(1):637--646, 2018.

\bibitem{reutergraphics}
{Reuters Graphics}.
\newblock Reuter graphics.
\newblock \url{https://graphics.reuters.com/}, 2019.
\newblock Accessed: 2019-10-04.

\bibitem{satyanarayan2017vega}
A.~Satyanarayan, D.~Moritz, K.~Wongsuphasawat, and J.~Heer.
\newblock Vega-lite: A grammar of interactive graphics.
\newblock {\em IEEE Transactions on Visualization and Computer Graphics},
  23(1):341--350, Jan 2017. doi: {{%
10\hspace{.1pt}\discretionary{.}{%
}{.}\hspace{.4pt}1109\discretionary{/}{%
}{/}TVCG\hspace{.1pt}\discretionary{.}{%
}{.}\hspace{.4pt}2016\hspace{.1pt}\discretionary{.}{%
}{.}\hspace{.4pt}2599030}}


\bibitem{savva2011revision}
M.~Savva, N.~Kong, A.~Chhajta, L.~Fei-Fei, M.~Agrawala, and J.~Heer.
\newblock Revision: Automated classification, analysis and redesign of chart
  images.
\newblock In {\em Proceedings of the 24th Annual ACM Symposium on User
  Interface Software and Technology}, UIST '11, pp. 393--402. ACM, New York,
  NY, USA, 2011. doi: {{%
10\hspace{.1pt}\discretionary{.}{%
}{.}\hspace{.4pt}1145\discretionary{/}{%
}{/}2047196\hspace{.1pt}\discretionary{.}{%
}{.}\hspace{.4pt}2047247}}


\bibitem{segel2011storytelling}
E.~Segel and J.~Heer.
\newblock Narrative visualization: Telling stories with data.
\newblock {\em IEEE transactions on visualization and computer graphics},
  16:1139--48, 01 2011. doi: {{%
10\hspace{.1pt}\discretionary{.}{%
}{.}\hspace{.4pt}1109\discretionary{/}{%
}{/}TVCG\hspace{.1pt}\discretionary{.}{%
}{.}\hspace{.4pt}2010\hspace{.1pt}\discretionary{.}{%
}{.}\hspace{.4pt}179}}


\bibitem{siegel2016figureseer}
N.~Siegel, Z.~Horvitz, R.~Levin, S.~Divvala, and A.~Farhadi.
\newblock Figureseer: Parsing result-figures in research papers.
\newblock In B.~Leibe, J.~Matas, N.~Sebe, and M.~Welling, eds., {\em Computer
  Vision -- ECCV 2016}, pp. 664--680. Springer International Publishing, Cham,
  2016.

\bibitem{sweller2011split}
J.~Sweller, P.~Ayres, and S.~Kalyuga.
\newblock The split-attention effect.
\newblock In {\em Cognitive load theory}, pp. 111--128. Springer, 2011.

\end{thebibliography}
